\documentclass[useAMS,usenatbib,12pt]{mn2e}
\usepackage[dvipdf]{graphicx,color}
\usepackage[latin1]{inputenc}
\usepackage{graphics}
\usepackage{amsfonts}
\usepackage{amsmath}
\usepackage{multicol}
\usepackage{layout}
\usepackage{amssymb}
\usepackage{epsfig}
\usepackage[a4paper]{hyperref}
\DeclareGraphicsExtensions{.eps}

\title[Halo creation in moving barrier models]{Dark matter halo creation in moving barrier models}
\author[J. Moreno et. al.]{Jorge Moreno$^{1}$, Carlo Giocoli$^{2}$ \& Ravi K. Sheth$^{1}$ 
\thanks{Email:
\href{mailto:jmoreno,shethrk@physics.upenn.edu}{jmoreno,shethrk@physics.upenn.edu},\,\href{mailto:carlogiocoli@unipd.it}{carlo.giocoli@unipd.it}}\\
$^{1}$Department of Physics and Astronomy, University of Pennsylvania, 209 South 33rd Street, Philadelphia, PA 19104-6396, USA\\
$^{2}$Dipartimento di Astronomia, Universita
degli Studi di Padova, Vicolo dell'osservatorio 2 I-35122 Padova, Italy}

\begin{document}
\date{}
\pagerange{\pageref{firstpage}--
\pageref{lastpage}} \pubyear{2009}
\maketitle
\label{firstpage}


\begin{abstract}
In hierarchical models of structure formation, the time derivative of the halo mass function may be thought of as the difference of two terms - a {\it creation} term, which describes the increase in the number of haloes of mass $m$ from mergers of less massive objects, and a {\it destruction} term, which describes the decrease in the number of $m$-haloes as these merge with other haloes, creating more massive haloes as a result.  The first part of this paper focuses on estimating the distribution of times when these creation events take place.  In models where haloes form from a spherical collapse, this distribution can be estimated from the same formalism which is used to estimate halo abundances: the constant-barrier excursion-set approach.  In the excursion set approach, moving, rather than constant-barriers, are necessary for estimating halo abundances when the collapse is triaxial.  First we generalise the excursion-set estimate of the creation time distribution by incorporating ellipsoidal collapse.  Then we show that these moving-barrier based predictions are in better agreement with measurements in numerical simulations than are the corresponding predictions of the spherical collapse model.  In the second part of the paper we link the creation times distribution to the creation term mentioned above.  For this quantity, the improvement provided by the ellipsoidal collapse model is more evident.  These results should be useful for studies of merger-driven star-formation rates and AGN activity.  We also present a similar study of creation of haloes conditioned on belonging to an object of a certain mass today, and reach similar conclusions - the moving barrier based estimates are in substantially better agreement with the simulations.  This part of the study may be useful for understanding the tendency for the oldest stars to exist in the most massive objects, and for star formation to only occur in lower mass objects at late times.  
\end{abstract}


\section{Introduction}

In hierarchical clustering models, self-bound dark matter haloes increase their mass by merging with other haloes \citep{press74,bond91,lacey93}. These mergers are expected to affect the galaxy populations hosted by the merging haloes \citep{white78,white91}, possibly triggering star formation or AGN activity \citep{efstathiou88,haehnelt93,kauffmann00}.  As a result, there has been some interest, both analytic and numerical, in estimating when such mergers happen.  

Let $n(m|t)\,{\rm d}m$ denote the number density of haloes with mass in the range d$m$ about $m$ at time $t$.  As a result of mergers, ${\rm d}n/{\rm d}t$ is the sum of two competing effects - the number of objects in a given mass bin increases if smaller mass objects merge to form an object of precisely this mass - an event we call {\it creation} - or the number decreases as objects of this mass merge with others, thus depleting the number in the bin - an event we call {\it destruction}.  Thus, the time derivative of the halo mass function is the difference of the creation and destruction rates: 
\begin{equation}
{\rm d}n/{\rm d}t = C(m,t) - D(m,t).
\label{coag}
\end{equation}
For the star formation and AGN problems above, one is more interested in the creation rates $C(m,t)$ than in ${\rm d}n/{\rm d}t$.  

Given $n(m|t)$, it is easy enough to take the time derivative; the problem is to separate ${\rm d}n/{\rm d}t$ into its two contributions.  Roughly speaking, low mass objects may have undergone significant mergers in the past but they are not being created in merging events any more - their evolution is expected to be dominated by the destruction term.  In contrast, extremely massive objects are undergoing substantial merging activity at the current time, and the time derivative of the halo mass function should be a good estimator of the creation rate of these objects.  But quantifying the general case requires a richer model.

Early work used the \cite{press74} form for $n(m|t)$, and advocated equating the `positive' term in ${\rm d}n/{\rm d}t$ with the creation rate, and the `negative' term with the destruction rate \cite[e.g.,][]{haehnelt98}.  But this is clearly not a solution at all, since it provides no rule for how to determine what one should correctly equate with `positive'.  For example, if ${\rm d}n/{\rm d}t = {\cal P}-{\cal N}$, there is no particular reason why one could not have written the right hand side as $({\cal P}-\epsilon) - ({\cal N}-\epsilon)$.   The second part of this paper is devoted to extracting the creation rate $C$ from ${\rm d}n/{\rm d}t$.  See \cite{blain93a,blain93b}, \cite{sasaki94} and \cite{kitayama96} for other attempts to solve this problem.

Before addressing the creation term $C$ (the number density of mergers per Gyr), in the first part of the paper we discuss the distribution of times $c(t|m)$ when these creation events take place (i.e., $c\equiv C/\int C\, {\rm d}t$ is the rate normalised by the total number of creation events that will ever occur).  Although $c$ and $C$ differ only by normalisation constant, it turns out that $c$ is somewhat easier to model.  This is because the excursion set formalism from which the Press-Schechter mass function can be derived \citep{bond91,lacey93}, carries with it a prescription for computing $c(t\,|m)$, the distribution of creation times \cite{percival99}.  In this case, the functional form of $c(t\,|m)$ is very similar to that of $f(m|t) \equiv m\,n(m|t)/\bar{\rho}$, where $\bar{\rho}$ is the mean comoving background density.  

Since that time, interest has shifted to functional forms for $n(m|t)$ which more closely approximate the abundances measured in numerical simulations \citep[e.g.,][]{sheth99,warren06,reed07,lukic07,tinker08}.  So it is interesting to ask how the creation time distributions are modified.  \cite{percival00} argue that the relation between the functional forms of $c(t|m)$ and $f(m|t)$ should survive these modifications, and show that this does indeed provide a good description of halo creation in simulations.  However, although they use intuition from the excursion set approach to motivate their arguments, their method side-steps the generalisation of the excursion set approach from which the modified mass functions may be derived -- this is the ellipsoidal collapse `moving' barrier approach \citep{sheth01,sheth02}.  The first goal of this paper is to calculate the creation time distribution self-consistently within the moving barrier excursion set approach.  We do find that $c$ and $f$ are simply related, but that this is actually extremely fortuitous -- Appendix~\ref{linB} demonstrates that this scaling does not hold generally.  

Getting the normalisation constant which relates $c$ to the creation rate $C(m,t)$ is a more challenging problem.  \cite{percival00} obtained this quantity by matching the creation time distribution to the rate measured in N-body simulations in the low redshift regime \cite[see also][]{percival99}, but they acknowledge that they have no theory for the normalisation factor.  One possible solution to this problem is to explore the evolution of the halo population in terms of coagulation theory, where the creation and destruction terms are estimated separately \citep{smoluchowski16,smoluchowski17}.  Early applications to galaxy formation and dark-matter halo interactions include \cite{silk78}, \cite{cavaliere91a}, \cite{cavaliere91b}, \cite{cavaliere92}, \cite{cavaliere93}, \cite{sheth97} and \cite{menci02}.  For a more recent treatment, see \cite{benson05} and \cite{benson08}.

For white-noise initial conditions, both the Smoluchowski and the Press-Schechter excursion-set expressions for $n(m|t)$ agree, so, for this case, the creation and destruction rates are known \citep{sheth97}.  However, obtaining the rates for more general initial conditions, or for the modified mass functions that are of more current interest, remains unsolved.  The second goal of the present paper is to provide a model for the creation rate of dark matter haloes that is informed by both coagulation theory and the modified excursion set approach with moving barriers.   

We also study the problem of how halo creation is modified if it is known that the merging haloes are bound up in objects of mass $M$ at some later time $T$.  The excursion set theory provides a way to compute the conditional mass function $N(m|t,M,T)$.  We show how the conditional distribution of creation times $c(t|m,M,T)$ is related to $f(m|t,M,T)=(m/M)N(m|t,M,T)$.  For the conditional rate, the problem is to separate ${\rm d}N/{\rm d}t$ into creation and destruction components.  \cite{sheth03} argues that this conditional distribution may be the basis for understanding the phenomenon known as down-sizing \cite[also see][]{neistein06}.  

Section~\ref{walks} provides a brief summary of the excursion set approach and shows why the creation times associated with different barriers differ.  Sections~\ref{times} and \ref{rates} compare the predicted and measured N-body creation time distributions and creation rates, respectively.  Conditional versions of these quantities are included in the appropriate sections.  A final section summarises our findings.  Technical details of the calculations are provided in the Appendices.  

A few final remarks regarding the different uses of the term `halo creation' are in order.  The first part of the paper focuses on the {\it creation time distribution}, $c(t|m)$, which can be derived within the excursion-set formalism.  The second part focuses on the {\it creation rate}, $C(m,t)$, the first term in the coagulation equation (equation~\ref{coag}).  The former is a {\it normalised} time distribution, while the latter is not normalised. (The normalised distribution is denoted with lower case $c$, while the un-normalised rate is denoted with capital $C$.)  Another source of confusion is that halo `creation' is distinct from halo `formation'; following \cite{lacey93}, the latter is typically defined as the time that an object first reaches half its current mass.  See \cite{giocoli07} for an explicit calculation showing how creation and formation are related.

\begin{figure*}
 \centering
 \includegraphics[width=\columnwidth]{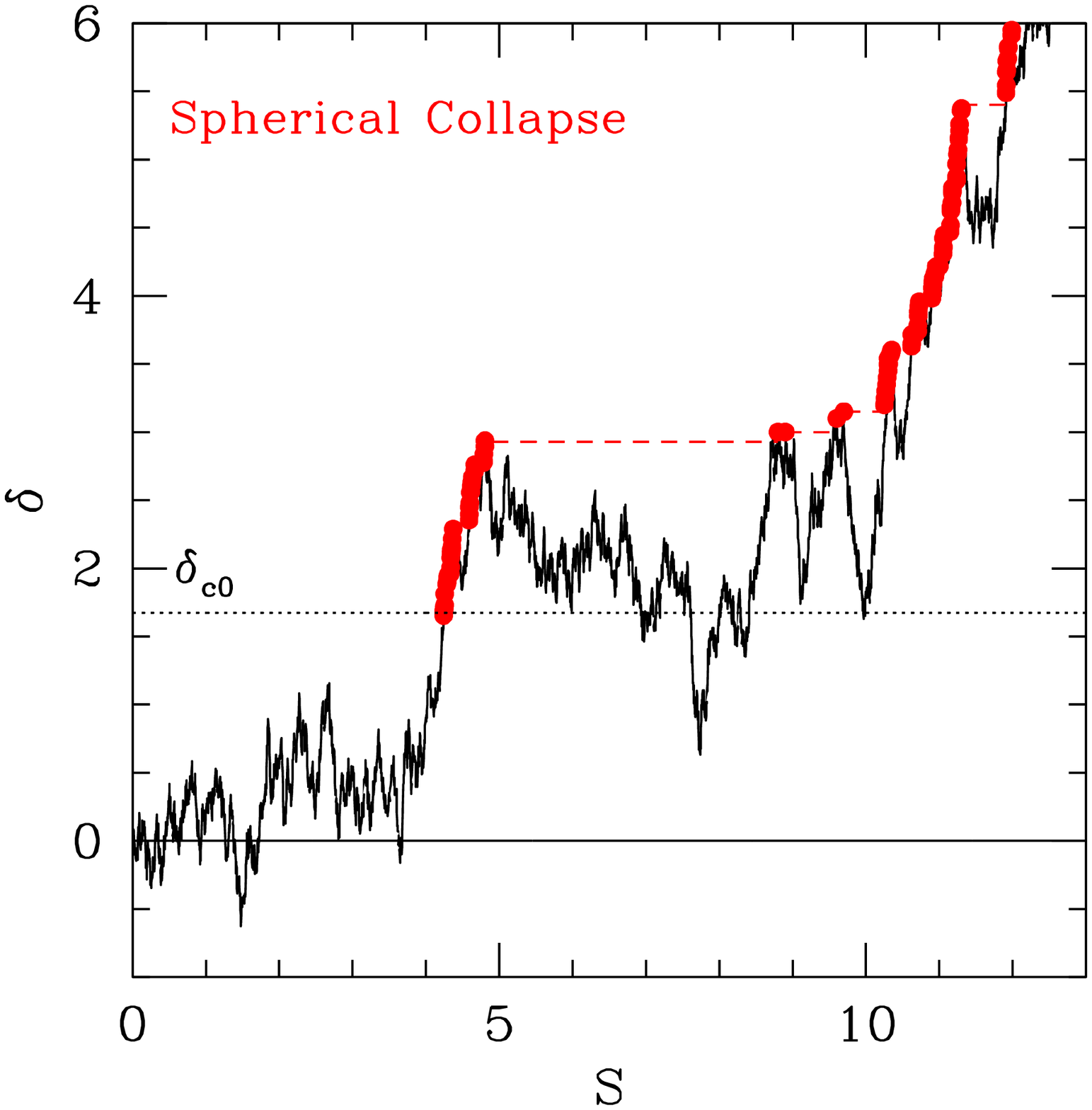}
 \includegraphics[width=\columnwidth]{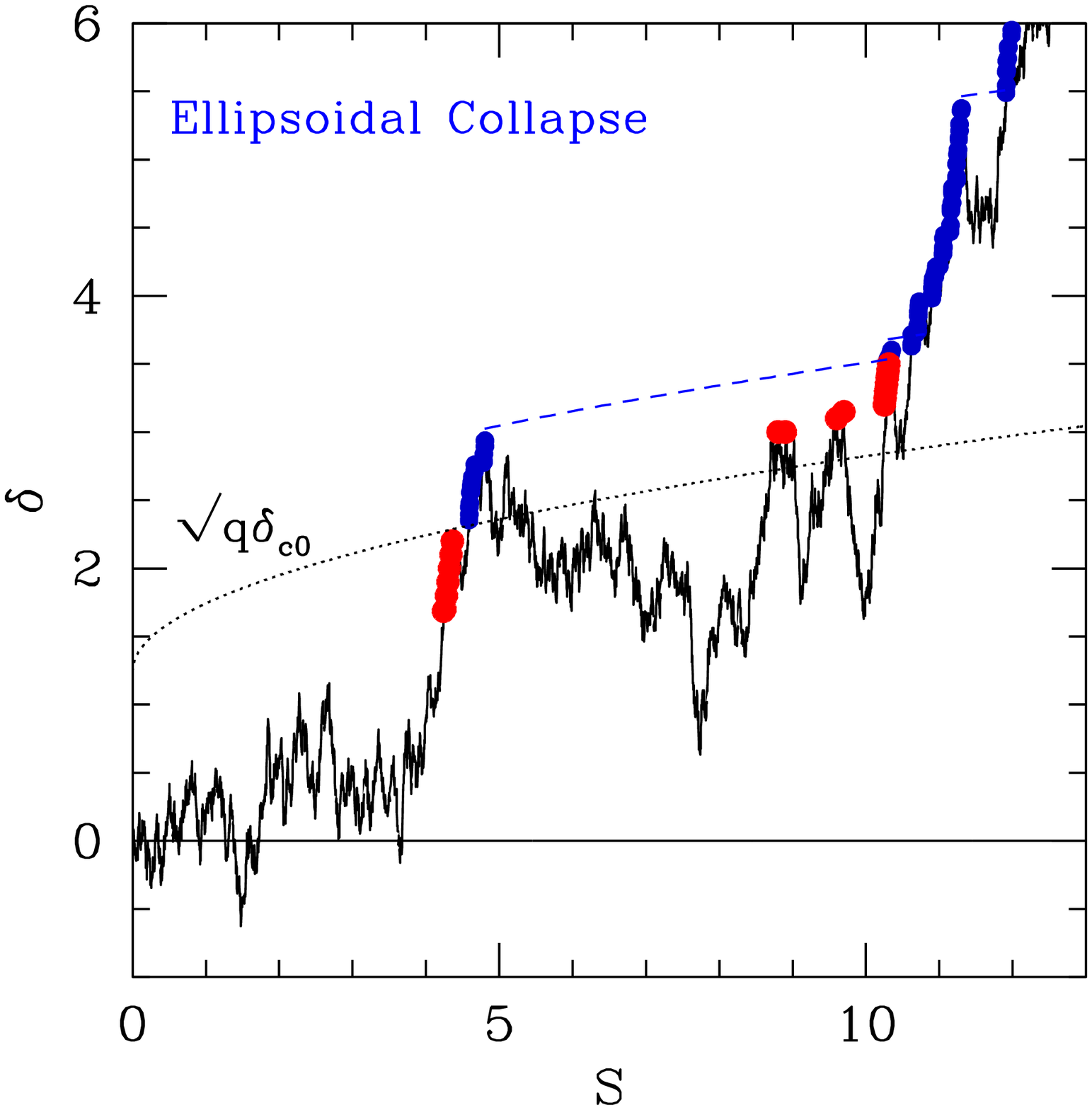}
 \caption{The mass history associated with a random walk (jagged line) depends on the properties of the barrier.  Panels show constant (left) and square-root (right) barriers.  Time increases as $\delta$ decreases and mass decreases as $S$ increases.  The filled circles on the random walk denote the history (red for spherical and blue for ellipsoidal collapse), and the horizonal jumps denote mergers.  In the right panel, the red circles were kept to emphasize that different barriers predict different mass histories for a given random walk.  For reference, the horizontal dotted line (curve) denotes the barrier associated with the present.}
 \label{history}
\end{figure*}

\section{EXCURSION SETS AND MASS HISTORY}\label{walks}

In the excursion set approach, the problem of estimating the halo abundances is mapped to one of estimating the distribution of the number of steps a Brownian-motion random walk must take before it first crosses a barrier of specified height \citep{bond91}.  In this approach, the height of the barrier plays a crucial role.  The Press-Schechter mass function is associated with barriers of constant height - such barriers arise naturally in models in which haloes form from a spherical collapse model. In constrast, the more accurate mass functions may be related to barriers whose height increases monotonically with the number of steps - such barriers arise naturally in ellipsoidal collapse models \citep{sheth01}.

Following \cite{sheth02}, we will be interested in barriers of the form 
\begin{equation}
\label{ellbarrier}
 B(S,\delta_{\rm c})  =\sqrt{q}\delta_{\rm c}\,
 \left\{ 1+\beta \left[\frac{S}{q \delta_{\rm c}^{2}}\right]^{\gamma}\right\}.
\end{equation}  
Here $\delta_{\rm c}$ is the overdensity required for spherical collapse - it is a monotonically decreasing function of time, given by $\delta_{\rm c0}/D(t)$, where $\delta_{\rm c0} \simeq 1.686$ and $D(t)$ is the growth factor.  $S$ is a monotonically decreasing function of halo mass, given by  $\sigma^{2}(m)$, the variance of the initial density fluctuation field. 
The Press-Schechter mass function is associated with $(q,\beta,\gamma)=(1,0,0)$, whereas ellipsoidal collapse has $(0.707,0.47,0.615)$.  

When $\beta=0$ (the barrier associated with spherical collapse) all walks are guaranteed to cross the barrier, and the barriers associated with two different times do not intersect.  However, if $\beta>0$ and $\gamma > 1/2$, e.g., for the ellipsoidal collapse barrier, then not all walks cross the barrier, and the barriers associated with two different times may intersect.  \cite{sheth02} suggest that the intersection of barriers may allow one to represent the possibility that haloes can fragment.  This is problematic when discussing the creation rate problem, which assumes that fragmentation never occurs.  

\begin{figure*}
 \centering
 \includegraphics[width=0.8\hsize]{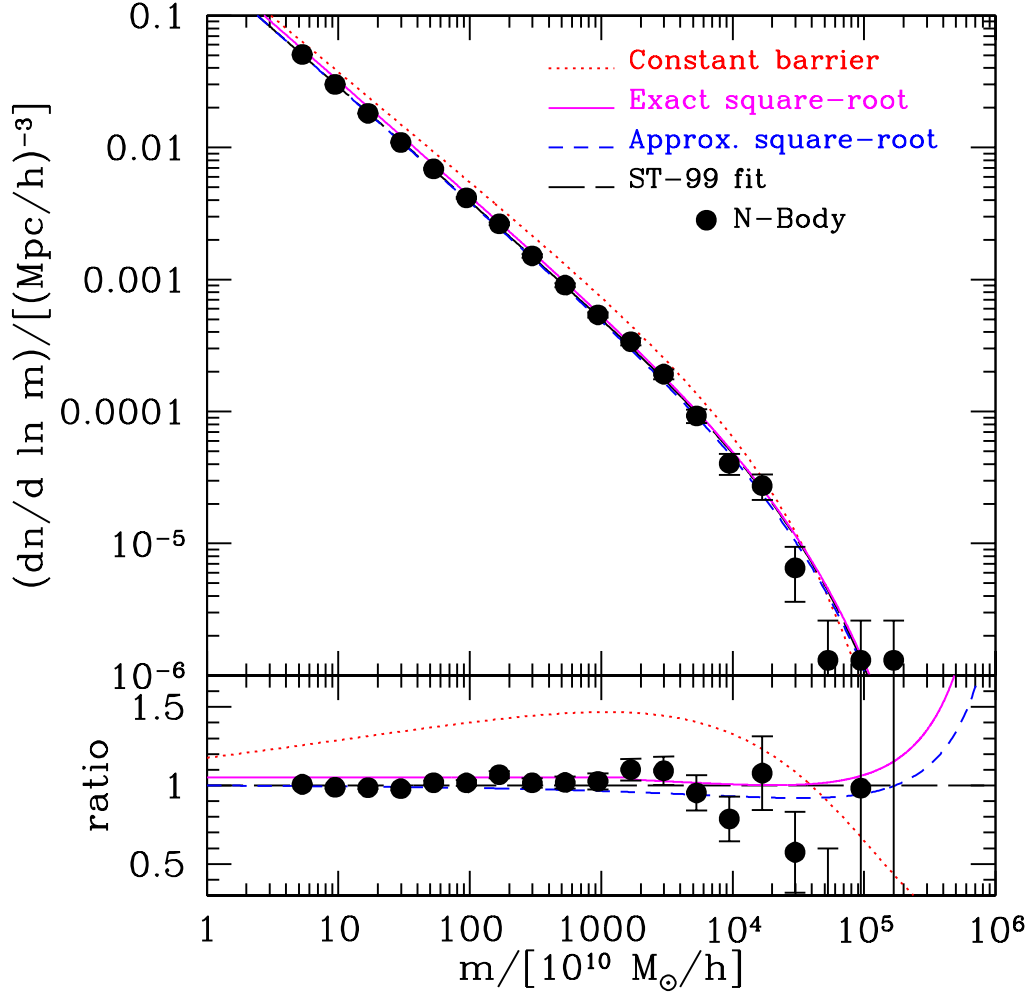}
 \caption{Comparison of the GIF2 mass function (symbols) with that derived from the first crossing distribution of a square-root barrier: equation~(\ref{sqrtbarrier}) with ($q,\beta,\gamma$) = (0.55,0.5,0.5).  The constant barrier Press-Schechter result is plotted for completeness.  Bottom panel shows the ratio of both data and theory curves to the functional form of \citet{sheth99}.}
 \label{massf}
\end{figure*}

For this reason, we study the limiting case of `square-root' barriers for which $\gamma = 1/2$:  
\begin{equation}
\label{sqrtbarrier}
 B(S,\delta_{\rm c}) = \sqrt{q}\delta_{\rm c} + \beta\sqrt{S}.
\end{equation}  
This family of barriers is particularly interesting because an analytic solution to the first crossing distribution is available \citep{breiman66}, although slightly cumbersome (see Appendix~\ref{ucondapp}).  
Because $\gamma=1/2$ is not very different from the value associated with ellipsoidal collapse, one might have expected the predicted halo abundances associated with square-root barriers to provide a reasonable description of simulations.  We show below that this can be achieved if one sets $(q,\beta,\gamma)=(0.55,0.5,0.5)$ (see Figure~\ref{massf}).  Halo merger and formation histories associated with this model are also in good agreement with simulations \citep{giocoli07,moreno08}.  

The dependence on $S$ of the square root barrier means that it is more like the ellipsoidal than spherical collapse barrier (which has constant height, independent of $S$).  However, there is one important respect in which the square root model is very like the constant one.  Consider the barriers associated with two different times.  For square-root barriers, the difference between the barrier heights is 
\begin{eqnarray}
 B(S,\delta_{\rm c2})-B(S,\delta_{\rm c1})
 &=&(\delta_{\rm c2}+\beta \sqrt{S})-(\delta_{\rm c1}+\beta \sqrt{S})
  \nonumber\\
 &=& \delta_{\rm c2}-\delta_{\rm c1}.
 \label{whyitworks}
\end{eqnarray}
Notice that this difference is independent of $S$.  This is also (trivially) true for constant barriers, but it is not true for any other values of $\gamma$.  In this respect, the excursion set model based on square-root barriers is extremely special.  This will be important later.

\subsection{Mass history for different barriers}

Figure~\ref{history} illustrates the relation between Brownian motion random walks and the mass growth history of an object.  Consider first the panel on the left.  The jagged line shows an example of a random walk --- this walk represents the run of smoothed overdensity around a randomly chosen position in the initial fluctuation field, as the region over which the overdensity is smoothed changes from large (left) to small (right).  (The plot actually shows the initial overdensity evolved to the present time using linear theory --- it differs from the initial overdensity by a multiplicative constant.)  The initial overdensities are all small compared to unity, so one may associate a mass with each smoothing scale: this mass is larger for the larger smoothing scales.  

Consider a horizontal line, and consider the places where it first intersects the random walk, as the height of this line, this barrier, is raised.  Clearly, this position shifts to the right as the barrier is raised (red filled circles) --- mass decreases as redshift increases.  Whereas the halo abundance problem corresponds to fixing the barrier height $\delta_{\rm c}$ (to illustrate, the height of dotted line corresponds to $\delta_{\rm c0}=1.686$) and asking for the distribution of $S$ values at which the barrier is first crossed, the halo creation problem corresponds to asking for the distribution of $\delta_{\rm c}$ values for a fixed $S$.  We will use  $f(S|\delta_{\rm c})\,{\rm d}S$ to denote the first crossing distribution, and $c(\delta_{\rm c}|S)\,{\rm d}\delta_{\rm c}$ to denote the distribution of creation times, where, for haloes of a given mass $m$, $c(\delta_{\rm c}|S)\,{\rm d}\delta_{\rm c} = c(t\,|m)\,{\rm d}t$.  

Notice that the mass increases relatively smoothly sometimes, and rather abruptly at others.  For instance, in the interval $4.2 \lesssim S \lesssim 4.8$ in the left panel, mass decreases smoothly.  Compare this situation to the sudden jump from $S\simeq 4.8 \rightarrow S \simeq 8.7$ (red long-dashed line).  Thus, this walk does not contribute to the creation time distribution for any values of $S$ between 4.8 and 8.3.  But it does contribute in the calculation of halo abundances for every $\delta_{\rm c}$.

It is interesting to compare this mass accretion history with that shown in the right panel.  The same walk is shown in both panels.  However, now the horizontal dotted line at $\delta_{\rm c0}$ has been replaced by a curve that has $\delta$-intercept at $\sqrt{q}\delta_{\rm c0}$, and increases with increasing $S$ --- this is the square-root barrier (equation~\ref{sqrtbarrier}) associated with the same epoch as the constant one.  The comparison clearly shows that the mass accretion history (blue solid circles) depends on the barrier shape.  For instance, the values with $S \lesssim 4.3$ and $S \simeq 8.8, 9.7$ and $S \lesssim 10.4$ are no longer included in the mass accretion history (red solid circles).  Moreover, even if a point on the random walk happens to be part of the mass history in both cases, its associated time is different under the two barrier prescriptions.  To illustrate, the point at $S \simeq 4.4$ is at the present for the square-root barrier case and in the!
  past when constant barriers are used.  As a result, the halo mass function and the creation-times distribution are modified.  One of the goals of this paper is to quantify these changes.

\begin{figure}
 \centering
 \includegraphics[width=\columnwidth]{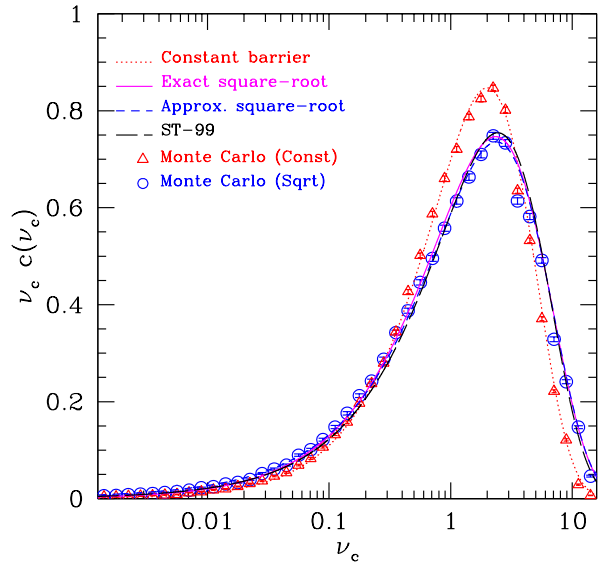}
 \caption{The creation time distribution in self-similar form.  The variable $\nu_{\rm c}$ denotes $\delta_{\rm c}^{2}/S$ at fixed mass.  Triangles and circles show the distribution measured from an ensemble of random walks with constant and square-root barriers respectively.  Dotted and solid (dashed) lines show the associated predictions.  The long dashed curve shows the result of inserting the \citet{sheth99} form for $f(\nu){\rm d}\nu$ into equation~(\ref{cvfv}).}
 \label{sswalk}
\end{figure}

\subsection{Self-similarity}
Barriers of the form given in equation~(\ref{ellbarrier}) are self-similar, in the sense that if $\delta_{\rm c}$ is increased by a factor $\kappa$, then so is $\sqrt{S}$.  As a result, the first crossing distribution $f(S|\delta_{\rm c})$ and the distribution of halo creation times $c(\delta_{\rm c}|S)$ are both simply functions of $\delta_{\rm c}^2/S$.  In the present paper we will denote this variable as $\nu$ if $\delta_{\rm c}$ is fixed and as $\nu_{\rm c}$ if $S$ is fixed.  In other words, if equation~(\ref{ellbarrier}) describes ellipsoidal collapse, then $f(S|\delta_{\rm c}){\rm d}S = f(\nu){\rm d}\nu$ and $c(\delta_{\rm c}|S){\rm d}\delta_{\rm c}=c(\nu_{\rm c}){\rm d}\nu_{\rm c}$.  Appendix~\ref{ucondapp} shows that 
\begin{equation}
\label{cvfv}
 c(\nu_{\rm c})\,{\rm d}\nu_{\rm c} = {\cal A}\sqrt{\nu_{\rm c}}\,f(\nu_{\rm c})\,{\rm d}\nu_{\rm c} 
\end{equation}
for the constant (${\cal A}=\sqrt{\pi/2}$) and square root (${\cal A} \simeq 2$) barriers.  This simple relationship is one of the central results of this paper, as is the warning that {it does not hold in general}.  Appendix~A uses a system of linear barriers to illustrate when this simple result does {\em not} apply.

To test equation~(\ref{cvfv}) for constant and square-root barriers, we generated $10^5$ random walks with $10^4$ steps between $S = 0$ and $S = S(m_{\rm p}) \simeq 28$.  Then we stored the corresponding mass histories for the constant and square-root barriers (e.g., solid circles in Figure~\ref{history}).  Every $(S,\delta_{\rm c})$ point along the history has a corresponding $\nu_{\rm c}$.  To study how the creation time distribution depends on $S$, we could have chosen the subset of walks which have the correct value of $S$, and then found the distribution of $\nu_{\rm c} = \delta_{\rm c}^2/S$ values for those walks.  However, the self-similar scaling above means that $c(\nu_{\rm c})$ should be the same for all $S$.  As a result, there is no need to select a subset in $S$ before binning in $\nu_{\rm c}$.  If we simply bin up all the $\nu_{\rm c}$ values, whatever the associated values of $S$, then we can compare the result with the predicted $c(\nu_{\rm c})$.  

The symbols in Figure~\ref{sswalk} show the creation time distributions for the constant (triangles) and square-root (circles) barriers:  the distribution associated with the square-root barrier is broader and peaks at slightly higher $\nu_{\rm c}$.  The curves show the predicted creation time distributions (equation~\ref{cvfv}); they are in excellent agreement with the measurements. Equation~(\ref{cvfv}) shows that these creation time distributions depend on the functional form of the first crossing distribution $f(\nu)$.  For the case of square-root barriers, we show the \cite{breiman66} exact but complicated expression for $f(\nu)$, and a much simpler approximation for it from \cite{sheth02}.  The two curves are almost indistinguishable.

The use of barriers which scale self-similarly (equation~\ref{ellbarrier}) was motivated by the observation that, when expressed as a function of $\nu$, halo abundances in simulations could be scaled to a universal form \citep{sheth99}.  We have added the long-dashed curve in the Figure, which shows the result of using the \cite{sheth99} functional form for $f(\nu_{\rm c})$ in equation~(\ref{cvfv}); it is almost indistinguishable from the curves associated with the square-root barrier.

\section{CREATION TIME DISTRIBUTION}\label{times}

In Section~\ref{walks} we discussed the creation time distribution from the excursion set point of view.  Having shown that the analytic expressions accurately reproduce our Monte Carlo measurements, we now study if they provide a good description of halo creation in N-body simulations.  

We use data from the GIF2 simulation \citep{gao04}, available at \href{http://www.mpa-garching.mpg.de/Virgo} {\texttt{http://www.mpa-garching.mpg.de/Virgo}}, which followed the evolution of $400^3$ particles of mass $m_{\rm p}=1.73 \times 10^9 h^{-1}\,M_\odot$ in a box of size of $110h^{-1}$Mpc, in a flat ${\rm \Lambda}$CDM cosmology with parameters ($\Omega_m,\sigma_8,h,\Omega_bh^2) = (0.3,0.9,0.7,0.0196)$.  Haloes were identified at 50 outputs equally spaced in $\log_{10}(1+z)$ between $1+z=20$ and $1+z=1$.  See \cite{giocoli08} for more details about the post-processing of the simulation.


Haloes were labelled as having been `created' if at least half of their particles were not observed in a more massive halo at an earlier time.  This is essentially the method adopted by \cite{percival99} -- and we refer the interested reader to that paper for details.   

\begin{figure*}
 \centering
 \includegraphics[width=\hsize]{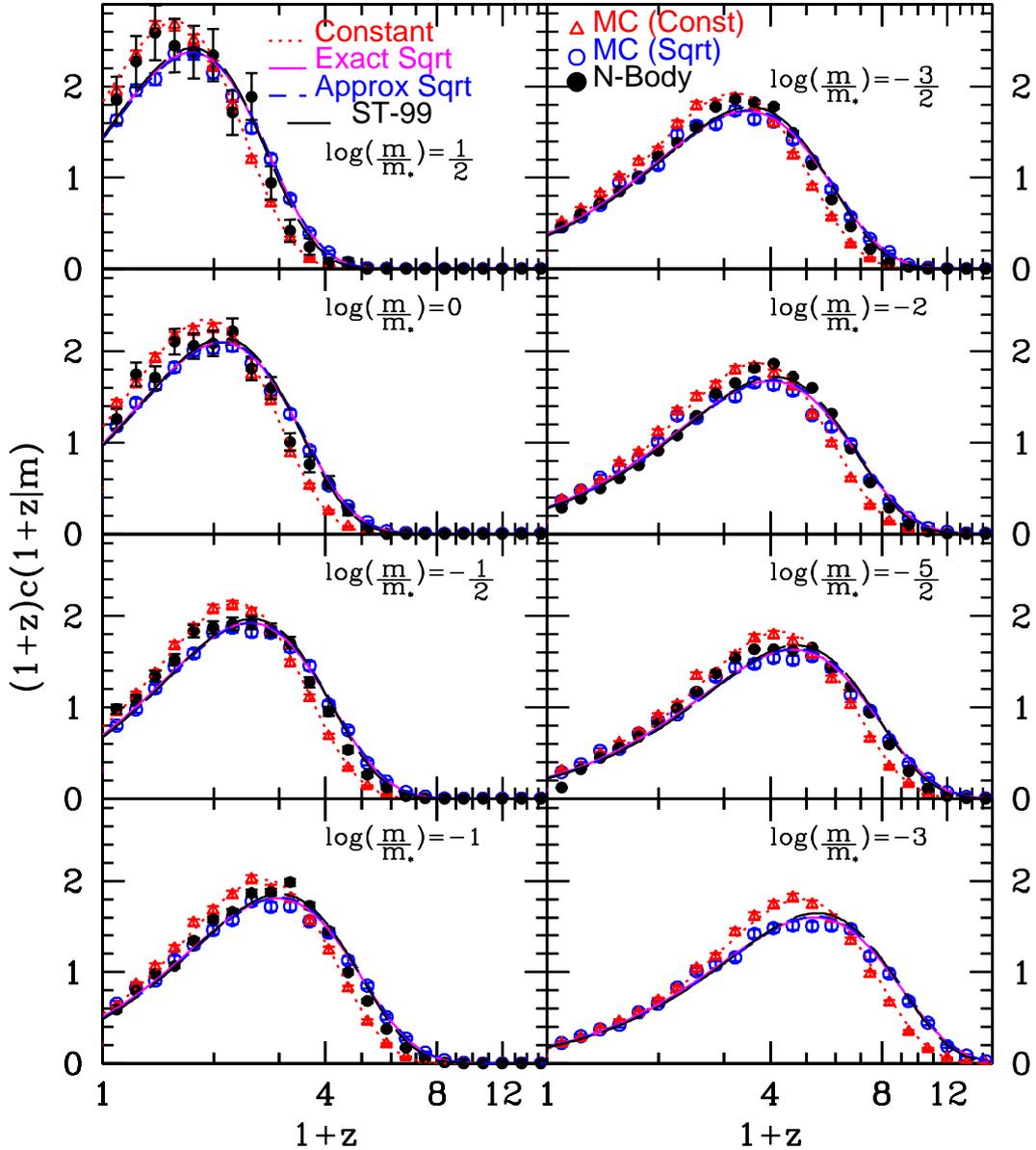}
 \caption{Distribution of creation redshifts for a number of bins in halo mass.  Filled circles show measurements in the simulations, and open triangles and circles show analogous measurements made from sampling our constant and square-root barrier random walk ensembles similarly to the simulations.  Dotted curve shows the prediction associated with a constant barrier; the exact square-root solution and its series approximation are the solid and dashed curves; long-dashed curve shows the result of inserting the \citet{sheth99} form into equation~(\ref{cvfv}).}
 \label{utimes}
\end{figure*}

\begin{figure*}
 \centering
 \includegraphics[width=\hsize]{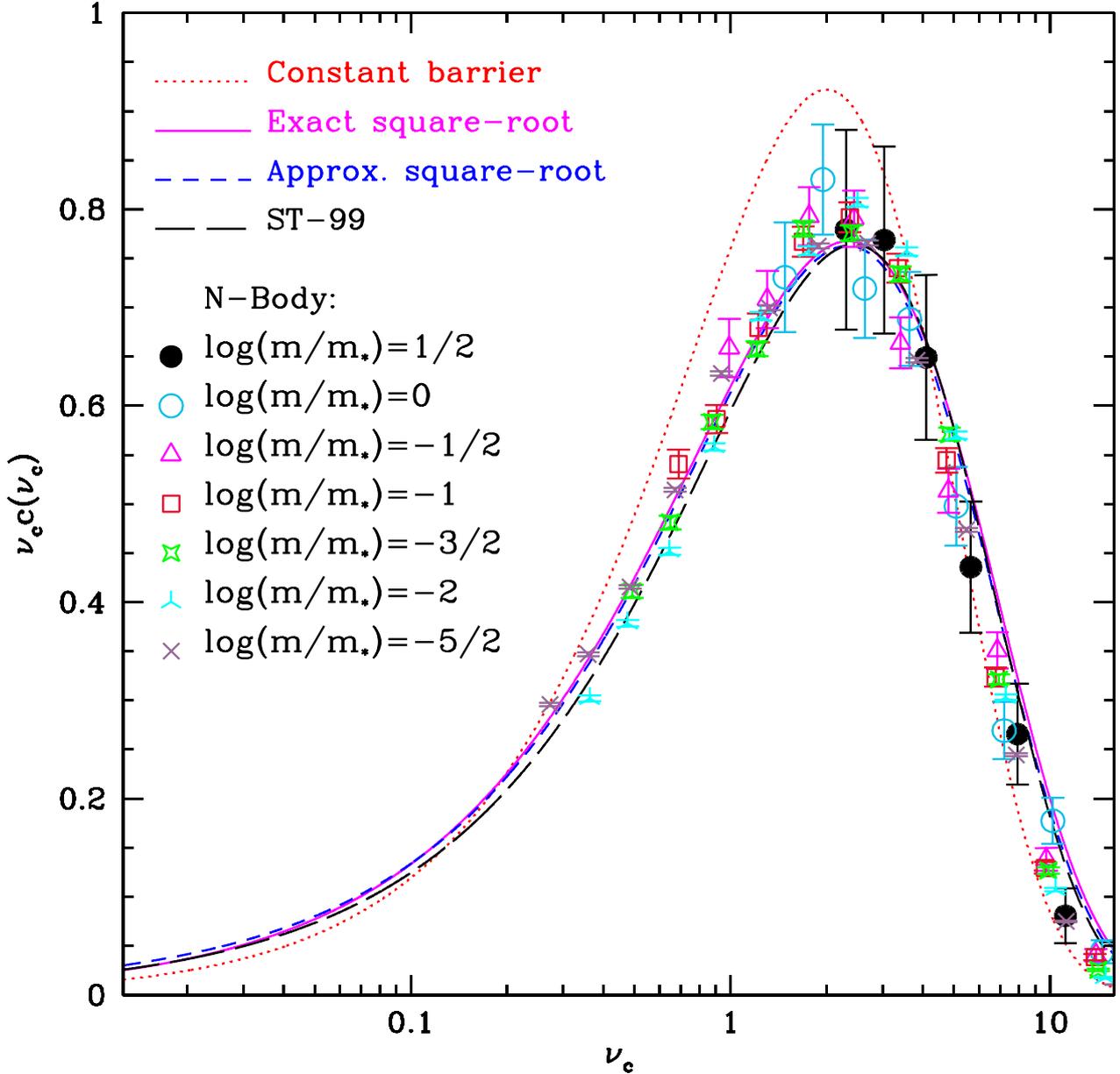}
 \caption{Universality of the distribution of halo creation times.  The same theory curves as in Figure~\ref{sswalk} are shown.   Different symbols show results for different halo masses in the GIF2 simulation data, as indicated.}
 \label{ssnbody}
\end{figure*}

\begin{figure*}
 \centering
 \includegraphics[width=\hsize]{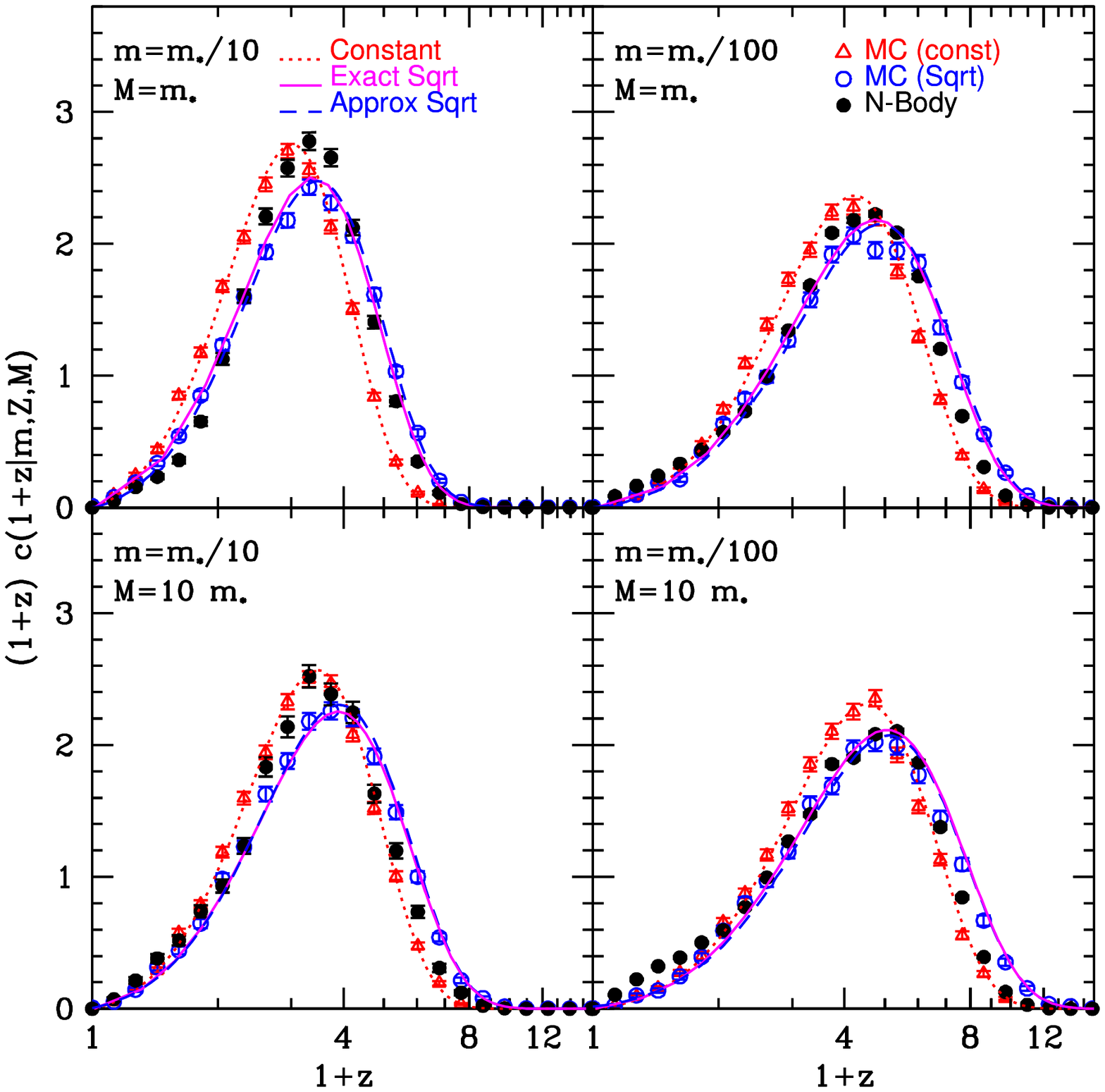}
 \caption{Conditional distribution of creation redshifts.  Symbols and style as in Figure~\ref{utimes}.  We plot $m=m_{\star}/10$ (left panels) and $m=m_{\star}/100$ (right panels) conditioned to end up in haloes of mass $M=M_{\star}$ (top panels) and $M=10 m_{\star}$ (bottom panels).  In all cases, $T$ denotes the present time.}
 \label{ctimes}
\end{figure*}

\subsection{Distribution of creation redshifts}

The creation time distributions we measure in simulations and shown as filled circles in Figure~\ref{utimes} are normalised to unity.  However, the simulations only sample $\delta_{\rm c}$ at epochs before the present time, whereas the theory curves assume that $0\le \delta_{\rm c}\le \infty$.  Therefore, for haloes of mass $m$, we set 
\begin{equation}
 c(z|m)\,{\rm d}z = \frac{c(\delta_{\rm c}|S_m)}{\int^{\infty}_{\delta_{c0}} {\rm d}\delta_{\rm c}' \, c(\delta_{\rm c}'|S_m) }
       \left|\frac{{\rm d}\delta_{\rm c}}{{\rm d}z}\right|\,{\rm d}z,
       \quad (\delta_{\rm c} \geq \delta_{\rm c0}).
\end{equation}
These are the curves in Figure~\ref{utimes}.  The open triangles and circles show the creation times measured in constant and square-root random walk ensembles sampled at the same redshifts as the simulations.  We studied bins of size d$\log_{10} m=0.2$ in mass centred at $\log_{10}(m/m_{\star}) = 0.5$ to $-3$ in steps of $-0.5$.  It is common practice to express halo masses in terms of the typical mass $m_{\star}(z)$, defined by $S(m_{\star}(z)) \equiv \delta_{\rm c}^{2}(z)$.  Throughout this work, $m_{\star}$ (with no $z$-dependence) denotes $m_{\star}(z=0)$.  In this cosmology, $m_{\star} =8.7 \times 10^{12} M_{\odot}h^{-1} \simeq 5030 m_{\rm p}$.  

A couple of remarks are in order.  Only the haloes with the highest redshift in each mass bin were treated as newly-created.  These measurement were tested with different bin sizes (not shown), yielding similar results.  One limitation is that if d$\log_{10} m$ is too small, most bins are empty, and the data does not follow a continuous curve.  One should not take d$\log_{10} m$ to be too large -- in particular, the mass bins for the different values of $m$ should be disjoint.  Our choice of d$\log_{10} m$ satisfied both criteria.  Lastly, we found that our choice of redshift bin d$\log_{10}(1+z)=0.05$ was sufficiently large to capture enough creation events and sufficiently small for comparison with the different theory curves.

The solid circles in the Figure denote the N-body measurement, where only haloes with $m>10 m_{\rm p}$ are considered (notice that the lowest-mass panel has no black filled circles).  In all cases, improvement over the location of the peaks is seen when the square-root barrier is used.  However, these curves are slightly broader than those traced by the simulation data, making the height of these normalised distributions lower.

\subsection{Self-similarity of halo-creation}\label{univ}

The excursion set model suggests that if the halo mass function $f(\nu){\rm d}\nu$ can be scaled to a self-similar form, then the creation time distribution $c(\nu_{\rm c}){\rm d}\nu_{\rm c}$ is also self-similar.  To test this we have scaled the values of $\delta_{\rm c}(z)$ associated with each mass bin in the simulations to $\nu_{\rm c}=\delta_{\rm c}(z)^2/S(m)$, and measured the resulting distribution of $\nu_{\rm c}$.  However, because all mass bins sample the same range in $\delta_{\rm c}$, they sample different ranges in $\nu_{\rm c}$.  We account for this by dividing the measured distribution of $\nu_{\rm c}$ by a normalisation factor given by $\int^{\infty}_{\nu_{\rm c0}} {\rm d}\nu_{\rm c}' \, c(\nu_{\rm c}')$, where $\nu_{\rm c0}=\delta_{\rm c0}^2/S(m)$ and $c(\nu_{\rm c})$ is associated with the \cite{sheth99} formula.

Figure~\ref{ssnbody} shows the result: different symbols show the rescaled distributions associated with the various masses.  Note that they do indeed appear to trace out a universal curve.  The various smooth curves show the constant barrier, square-root barrier, and \cite{sheth99} based predictions.  The symbols approximately split the difference between the constant and square-root barrier models.

\subsection{Conditional distribution of creation redshifts}

So far we have discussed the unconditional creation-time distribution $c(t\,|m)$ and its relation to $f(m|t)$.  In this section we study the creation time distribution, $c(t\,|m,T,M)$, of $m$-haloes at time $t$ conditioned to be bound up in $M$-haloes at a later time $T$.  We also discuss how this quantity is related to $f(m|t,M,T)$, the fraction of mass in $m$-progenitors at time $t$ of a final halo of mass $M$ at time $T$.  The latter is derived from the excursion-set theory by setting
\begin{equation}
\label{condf}
f(m|t,M,T){\rm d}m=f(S|\delta_{1},S_0,\delta_0){\rm d}S,
\end{equation}
where $S=S(m)$, $S_0=S(M)$, $\delta_{1}=\delta_{\rm c}(t)$ and $\delta_{0}=\delta_{\rm c}(T)$.  The right-hand term is the crossing distribution of a barrier $B(S,\delta_{1})$ by random walks with origin at $(S_0,B(S_0,\delta_{0}))$.  In Figure~\ref{history}, this amounts to shifting the origin from $(0,0)$ to $(S_0,\delta_0)$ (left panel) or to $(S_0,\sqrt{q}\delta_0+\beta\sqrt{S_0})$ (right panel).  In this work, the conditional mass function is denoted by $N(m|t,M,T)$d$m=(M/m)f(m|t,M,T)$d$m$.
 
In essence, conditioning is equivalent to finding the (unconditional) crossing distribution of a barrier
\begin{equation}
\label{dbarrier}
{\cal B}(s,\delta_{1},\delta_{0})=B(s+S_0,\delta_{1})-B(S_0,\delta_{0}),\,\,\,s=S-S_0.
\end{equation}
In the constant barrier problem, ${\cal B}=\delta_1-\delta_0$.  This means one simply replaces all the unconditional expressions given previously with $\delta_{\rm c}(z) \rightarrow \delta_{1}-\delta_{0}$ and $S \rightarrow s=S-S_{0}$.  As a result, the only change occurs in the self-similar variable $\nu$ (and $\nu_{\rm c}$):  $\delta_{\rm c}^2/S \rightarrow (\delta_{1}-\delta_{0})^{2}/(S-S_{0})$ \citep{lacey93,percival99,sheth03}.

The square-root barrier is slightly more complicated:
\begin{equation}
 {\cal B}(s,a_{\rm c})= a_{\rm c} + \beta \sqrt{s+S_{0}},\,\,\, a_{\rm c} = \delta_{1}-\delta_{0}-\beta \sqrt{S_0}.
 \label{bs}
\end{equation}
Because equation~(\ref{bs}) is not quite the same functional form as equation~(\ref{sqrtbarrier}), the first crossing distribution is not simply a suitably rescaled version of the unconditional distribution.  Rather, it is a function of $\eta_{\beta} \equiv a_{\rm c}/\sqrt{S_{0}}$ and $s/S_0$.  Nevertheless, the logic one follows to arrive at the creation time distribution is the same.  In particular, the conditional versions of $c$ and $f$ are simply related:
\begin{equation}
c(\eta_{\beta}|S/S_0){\rm d}\eta_{\beta}={\cal A}_{(S/S_0)} f(S/S_0|\eta_{\beta}){\rm d}(S/S_0),
\label{Acond}
\end{equation}
where now the ${\cal A}$ factor is a function of $S/S_{0}$.  This is another central result of this paper (compare to equation~\ref{cvfv}).  Expressions for the exact solution and the corresponding \cite{sheth02} approximation are given in Appendix~\ref{condapp}. 

Figure~\ref{ctimes} shows the conditional distribution of creation redshifts for $m/m_{\star}=(0.1,0.01)$ that end up in haloes of mass $M/m_{\star}=(1,10)$ today.  Filled symbols show the GIF2 measurements, open circles show our Monte Carlos with square root barrier, and open triangles show Monte Carlos with a constant barrier.   
Smooth curves show the corresponding predictions -- notice that they are in excellent agreement with the Monte Carlos.  

The simulation bin sizes d$\log_{10} m$ and d$\log_{10}(1+z)$ were used as in the unconditional case.  Selecting haloes to be conditioned to belong to a final $M$-halo reduces the number of creation events significantly.  We selected haloes bound to end-up in clumps with mass in a bin of size d$\log_{10} M=0.5$.
As in the unconditional case (Figure~\ref{utimes}), the distributions peak at higher redshifts and are slightly broader for lower $m$ (compare left and right panels).  The same trends are seen as one increases the final $M$ (compare top and bottom panels).

In general, the moving barrier based curves provide a much better description of the simulations, although the agreement is by no means perfect.  For example, the square root barrier tends to produce distributions which are slightly broader than those in the N-body simulation.  This effect is more pronounced in the right-hand side panels.  A similar effect was found by \cite{moreno08} for the formation time distribution, which is related to (but different from) the creation time distribution of interest here \cite[see][for details]{giocoli07}.  In that paper, we speculate that the discrepancy there was due to non-Markovian effects.  We refer the interested reader to \cite{pan08} for a discussion of this topic.

\begin{figure*}
 \centering
 \includegraphics[width=\hsize]{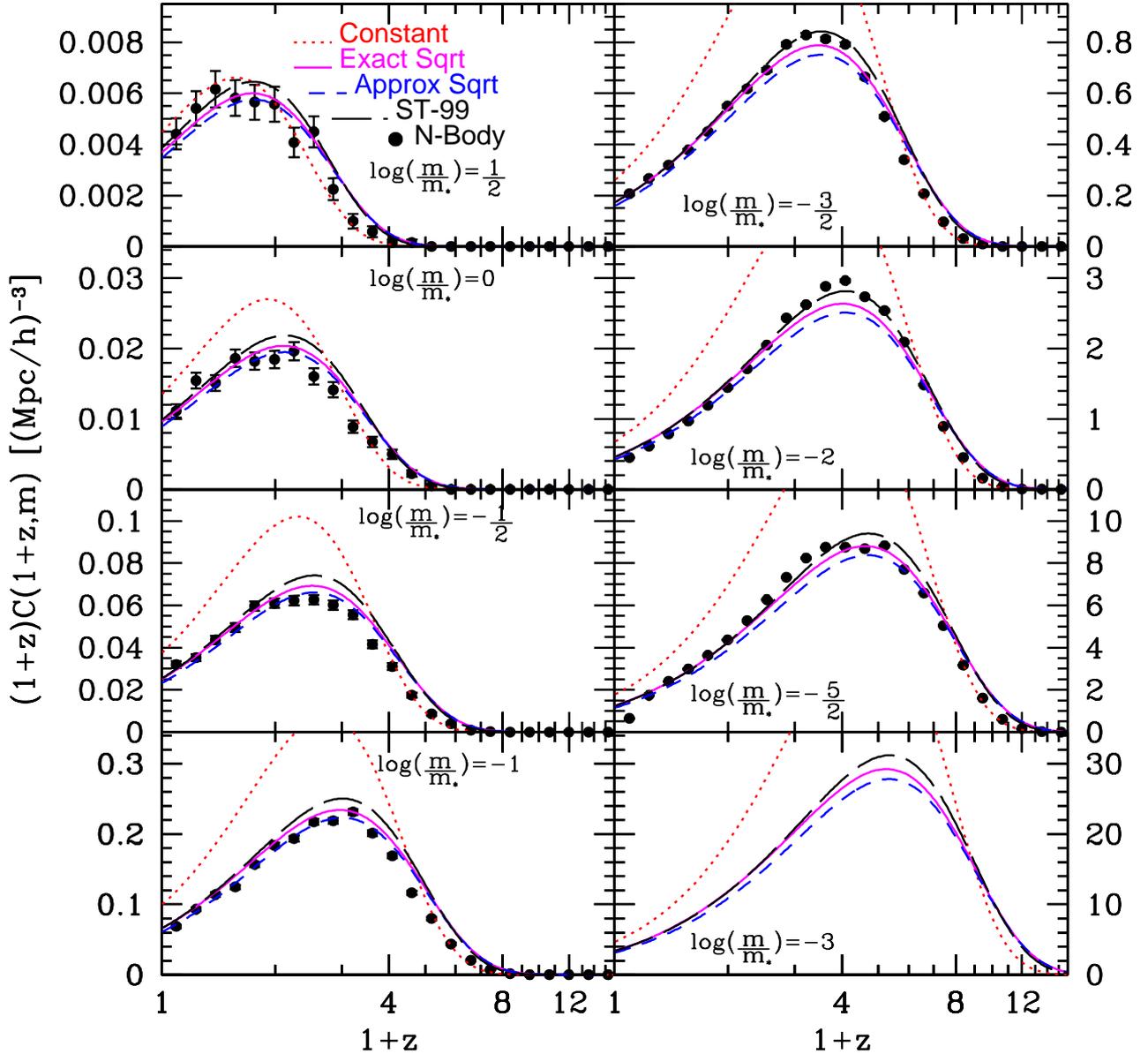}
 \caption{Halo creation rates.  Symbols, line styles and choices of mass as in Figure~\ref{utimes}.  As in Figure~\ref{utimes}, the lowest panel on the right shows not data because this case is beneath the resolution of the N-body simulation.}
 \label{urate}
\end{figure*}

\begin{figure*}
 \centering
 \includegraphics[width=\hsize]{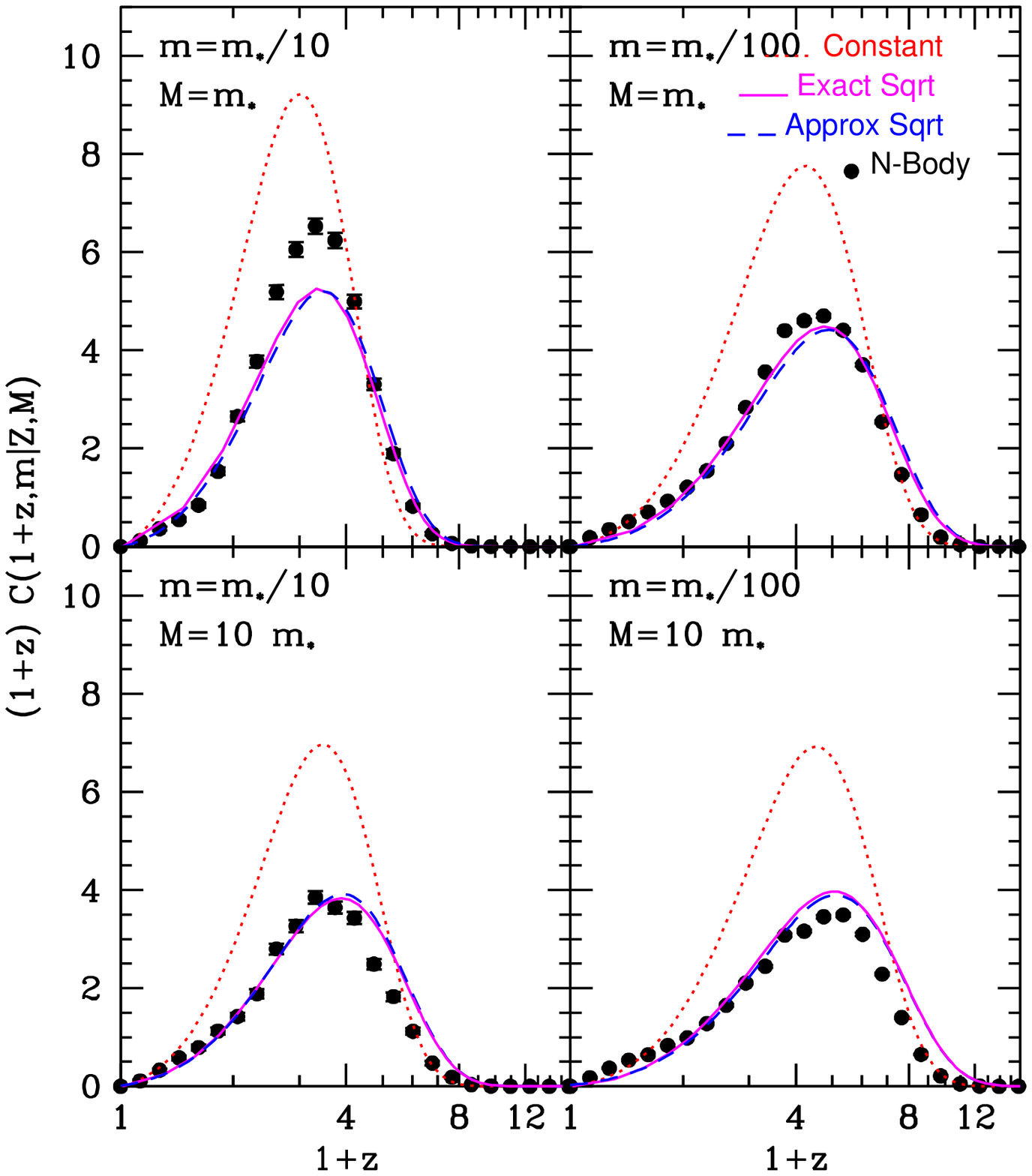}
 \caption{Conditional creation rates.  Symbols, line styles and choice of masses as in Figure~\ref{ctimes}.}
 \label{crate}
\end{figure*}

\section{HALO CREATION RATES}\label{rates}

Extracting the creation rate from ${\rm d}n/{\rm d}t$ (or ${\rm d}N/{\rm d}t$) is a non-trivial problem.  In this section we make use of halo coagulation theory to estimate this quantity.

\subsection{Unconditional rate}

In this formalism, the halo mass function $n(m|t)$ obeys 
\begin{equation}
 \frac{{\rm d}n(m|t)}{{\rm d}t}=C(m,t)-D(m,t),
\end{equation}
where 
\begin{equation}
 \label{csmol}
 C(m,t)=\int^{m}_{0}\frac{K(m',m-m';t)}{2}n(m'|t)n(m-m'|t){\rm d}m',
\end{equation}
is our creation term, and the destruction term is 
\begin{equation}
 \label{dsmol}
 D(m,t)=\int^{\infty}_{0}K(m,m';t)n(m|t)n(m'|t){\rm d}m'
\end{equation}
\citep{smoluchowski17}.
In these expressions, the coagulation kernel $K(m,m';t)$ is symmetric in $m$ and $m'$.  

Few analytic solutions to Smoluchowski's equation exist.  However, when the kernel is additive in mass, then the associated mass function is given by the Press-Schechter formula for white-noise initial conditions \citep{silk78,sheth97}.  Of course, white-noise is a bad approximation to the initial conditions in the CDM models of current interest.  Moreover, we have shown that ellipsoidal collapse gives a better description of halo abundances and creation times than Press-Schechter (spherical collapse).  Nevertheless, the expression obtained for the creation term in that special case,
\begin{equation}
\label{mn}
 C(m,t) = \bar{\rho}\, m\, n(m|t) 
          \Big{|} \frac{{\rm d}\delta_{\rm c}}{{\rm d}t}\Big{|}
\end{equation}
\citep{sheth97}, will serve as a guide (see Appendix~B for more details regarding coagulation with spherical collapse and white-noise initial conditions). 

This prescription has two interesting properties.  First, it is related to the creation time distribution in a simple way:
\begin{equation}
  C(m,t)=g(m)c(t|m)
\end{equation}
(please see Appendix~\ref{Cc}).  This is equivalent to saying that one obtains $c$ by normalising $C$: 
\begin{equation}
\frac{C(m,t)}{\int C(m,t) {\rm d}t}=\frac{g(m)c(t|m)}{\int g(m)c(t|m) {\rm d}t}=
\frac{g(m)c(t|m)}{g(m)\int c(m,t) {\rm d}t}=c(t|m).
\end{equation}
This indicates that all the time dependence in the creation rate is encoded in the creation time distribution, which was amply studied in the first part of the paper.  In particular, in Section~\ref{times} we showed that the mass function and the creation time distribution are related in a simple way (equation~\ref{cvfv}), at least for barriers which are close to constant or square-root.   Since we are in this regime, we assume that our prescription (i.e., $C(m,t)=g(m)c(t|m)$) works well enough for our purposes, even when the initial power spectrum is not white noise, and the barrier associated with the random walk problem which gives the mass function $n$ that is of interest, is not constant.  The second property is that this prescription reduces to the known exact creation rate in the white-noise case.  This fact, while seemingly obvious, was never imposed as a requirement to be obeyed by the creation rate in previous works \citep[][]{blain93a,blain93b,sasaki94,kitayama96}.

Figure~\ref{urate} compares this assumption with the measured creation rates in the simulations.  The simulation measurements in Figures~\ref{urate} and \ref{utimes} are the same, except that in the latter, data is normalised in time.  Notice that the heights of the curves increase with decreasing mass.  This reflects that fact that, in hierarchical models, more small haloes are created during the history of the Universe than are massive halos (which are only created at later times).  The constant barrier model works well for massive haloes, but it overpredicts the creation rate of less massive haloes -- showing a similar discrepancy as in the Press-Schechter mass function in that mass regime.  In all cases, the square-root barrier and the \cite{sheth99} creation rates match N-body results reasonably well.


\subsection{Conditional rate}

We assume that the same prescription can be applied for the conditional case.  I.e., the rate of creation of haloes of mass $m$ at time $t$ conditioned to belong to $M$ haloes at a later time $T>t$ is given by
\begin{equation}
\label{mN}
C(m,t\,|M,T)=\bar{\rho}\,m\,N(m,t\,|M,T)\,\Big{|} \frac{{\rm d}\delta_{\rm c}}{{\rm d}t} \Big{|}
\end{equation}
\citep{sheth03}.  Figure~\ref{crate} shows our results for the same set of $m$ and $M$ as in Figure~\ref{ctimes}.  As in the unconditional case, the simulation measurements in Figures~\ref{crate} and \ref{ctimes} are the same, except that in the latter, data is normalised in time.

For both choices of $m$, the height of the curves decreases with increasing final mass $M$ (compare top and bottom panels).  This effect is less pronounced for the smaller $m$ (compare left and right panels).  In all panels, curves based on the square-root barrier provide a more accurate description of the measurements.

\section{Discussion and Conclusions}\label{discuss}
We have used the excursion set approach to study how the distribution of halo creation times is modified if haloes are assumed to form from an ellipsoidal rather than a spherical collapse (see Figure~\ref{history}).  The creation time distribution governs the time dependence in the creation term found in Smoluchowski-like interpretations of the evolution of the halo mass function \citep{sheth97,benson05,benson08} (i.e., normalising the creation rate in time gives the creation time distribution).  In this approach, halo abundances and creation times can be derived from the study of random walks which cross barriers of some specified height.  The barrier itself is specified by the physics of gravitational collapse; that associated with spherical collapse has a constant height and this allows simple analytic solutions for halo abundances and creation times \cite[e.g.,][]{kitayama75,bond91,percival99}.  Dropping the spherical collapse assumption is not trivial, as it results in a barrier whose height increases with distance along the walk \citep{sheth01}.  Moveover, the moving barrier associated with ellipsoidal collapse is not well-suited to the study of creation times \cite[see equation~\ref{ellbarrier} and related discussion, as well as discussion in][]{sheth02}.  

For this reason, we approximated the ellipsoidal collapse barrier with a square-root barrier (equation~\ref{sqrtbarrier}) for which an analytic solution for the halo mass function exists \citep{breiman66}.  Moreover, this barrier yields a halo mass function which is in good agreement with simulations (Figure~\ref{massf}).  We used Monte-Carlo realisations of random walks to show that the associated creation time distribution is related to the halo mass function in just the same way that it is for the constant barrier (equation~\ref{cvfv}).  Because the halo mass functions associated with the two barriers are different, the creation time distributions also differ (Figure~\ref{sswalk}).  The moving barrier based predictions were in slightly better agreement with measurements of halo creation times in simulations (Figures~\ref{utimes} and~\ref{ssnbody}).  

We also presented the first derivation of the conditional creation time distribution using the moving-barrier model of ellipsoidal collapse.  In this case, the differences between the constant and moving barrier predictions were somewhat more dramatic; the moving barrier predictions were in substantially better agreement with the simulations (see Figure~\ref{ctimes} -- especially the panels on the right).  For both the unconditional and conditional versions we showed there was a simple link between $c$ and $f$ (equations~\ref{cvfv} and \ref{Acond}), but showed that this link is special to the constant and square-root barriers.  While not true in general (see Appendix~A for a counter example), we argued that, for the mass functions of current interest in cosmology, this link should provide reasonably accurate approximations to the creation time distribution.  

We also presented an approximation for the normalisation constant which converts the creation time distribution into a halo creation rate.  This was motivated by connecting our results to coagulation theory \cite[e.g.][]{sheth97}.  
Figures~\ref{urate} and \ref{crate} indicate that this yields reasonably accurate results.

An alternative approach is to express the creation rate in terms of $R(m,m'|t)$, the rate of mergers of $m$-haloes with $m'$-haloes, creating $(m+m')$-haloes as a result.  This merger rate has been measured recently in cosmological simulations \citep{fakhouri08a,fakhouri08b} and studied in the spherical \citep{lacey93} and ellipsoidal \citep{zhang08a} collapse versions of the excursion set approach.  The problem in this case is that, with the exception of white-noise initial conditions, the excursion set theory predicts that $R(m,m'|t) \not= R(m',m|t)$ \citep{sheth97} \cite[but see][for a possible solution]{neistein08}.  Even if the merger rate were defined unambiguously within the excursion set approach with moving barriers, care must be taken when the integrals in equations~(\ref{csmol}) and (\ref{dsmol}) are computed.  Our prescription of the creation rate circumvents these problems altogether.  A third possible method is to use the merger rate developed by \cite{benson05} \cite[see also][]{benson08}.  This method avoids the complications in the excursion set theory by estimating the coagulation kernel $K$ numerically.  Unfortuntely, the answer in this case depends on the choice of regularisation technique.  The formulation a complete model for halo mergers with general initial conditions and with the advantages of ellipsoidal collapse remains an open, and quite interesting, problem.

Halo creation rates have been used to model star formation and AGN activity.   We refer the reader to \cite{haehnelt93}, \cite{haehnelt98}, \cite{haiman98}, \cite{haiman00}, \cite{hosokawa02}, \cite{granato04}, \cite{lapi06}, and \cite{wang09} for applications of halo creation where our results can provide an improvement.  In some cases, the derivative of the mass function d$n/$d$z$ is used to replace the creation rate -- while in others, the creation rate is extracted from d$n/$d$z$ without proper justification (see the references in the Introduction).  It is our hope that our analysis will be useful for such studies.  We caution, however, that a halo may be created smoothly through a process of gradual accretion, or violently, through a merger.  If this difference matters, it may be more appropriate to use a full merger history tree.  \cite{moreno08} describes such moving-barrier based trees -- in this context, also see \cite{hiotelis06,parkinson07,neistein08} and \cite{zhang08b}.  

The phenomenon called `down-sizing' is often associated with two trends:  
(a) the tendency for the most massive galaxies to host the oldest stars, and 
(b) the tendency, at later times, for star formation to occur in haloes of lower mass.  
Our creation time distributions conditioned on final halo mass have been used to understand trend (a):  if star formation only occurs in sufficiently massive haloes, then (a) arises naturally in hierarchical models \citep{sheth03}.  This was confirmed by \cite{neistein06}, who then found that they were unable to explain trend (b) using this same mechanism.  However, a corrollary of Sheth's argument is that, if star formation does {\em not} occur in haloes {\em above} a certain critical mass, then trend (b) is the result.  That is to say, if stars only form in halos between a minimum and maximum mass range, then down-sizing trend (a) is the result of the minimum mass, and trend (b) is the result of the maximum mass.  Therefore, we expect our results to provide further insight into this phenomenon in general, and into the critical mass scales in particular.

\section*{acknowledgments}
We wish to thank R.~E.~Smith and G.~Tormen for clarifying our questions on Bayesian statistics and on N-body simulations, respectively.  J.~Moreno thanks A.~Benson for sharing his ideas on the relationship between halo mergers and polymer growth.  Also, we express our gratitude to F.~Shankar, D.~H.~Weinberg and J.~Hyde for other stimulating discussions and comments.  Lastly, we thank our anonymous referee for his or her input, which improved this work greatly.  This work was supported in part by CONACyT--Mexico, and by Grant 2002352 from the US-Israel BSF.

\bibliographystyle{mn2e}


\appendix

\section{EXCURSION SET RESULTS}

\subsection{Unconditional formulae}\label{ucondapp}

In this section we provide expressions for $f(\nu){\rm d}\nu$ and $c(\nu_{\rm c}){\rm d}\nu_{\rm c}$.
\begin{itemize}
\item Constant barrier (Press-Schechter):
\begin{equation}
f(\nu){\rm d}\nu = \sqrt{\frac{\nu}{2\pi}}{\rm e}^{-\frac{\nu}{2}}\frac{{\rm d}\nu}{\nu}\,.
\label{fnu}
\end{equation}
\item Square-root barrier (exact):
\begin{equation}
f(\nu){\rm d}\nu=\sum_{\{\lambda\}}\nu^{\frac{\lambda}{2}} l_{\lambda}(-\beta) \frac{{\rm d} \nu}{\nu}, \,\,l_{\lambda}(-\beta) \equiv \frac{{\rm e}^{-\beta^{2}/4}}{2}\frac{D'_{\lambda}(-\beta)}{I_{\lambda}(-\beta)}.
\label{mrnu}
\end{equation}
where $D_{\lambda}(x)$ are the parabolic cylinder functions, 
\begin{eqnarray}
D_{\lambda}'(x)=\frac{{\rm d}D_{\lambda}(x)}{{\rm d}x}, \,\,\,I_{\lambda}(-\beta)=\int^{\infty}_{-\beta}{\rm d}x\,D^{2}(x),
\end{eqnarray}
and the $\{\lambda\}$-eigenvalues satisfy $D_{\lambda}(-\beta)=0$ \citep{breiman66,mahmood05,giocoli07}.  
\item Square-root barrier (series approximation):
\begin{equation}
f(\nu){\rm d}\nu = \sqrt{\frac{\nu}{2 \pi}}{\rm e}^{-\frac{\nu}{2}[1+\frac{\beta}{\sqrt{\nu}}]^{2}} [1 +\frac{\beta \alpha}{\sqrt{\nu}}  ] \frac{ {\rm d}\nu}{\nu}, \,\,  \alpha \simeq 0.2461. 
\end{equation}
See \cite{sheth02}.
\item `Sheth-Tormen' result:
\begin{equation}
f(\nu){\rm d}\nu = A \sqrt{\frac{\nu}{2\pi}}{\rm e}^{-\frac{\nu}{2}}[1+\nu^{-p}]\frac{ {\rm d} \nu}{\nu},\,\,(A,p)=(0.332,0.3).
\label{stfit}
\end{equation}
See \cite{sheth99}. 

In the above models we have suppressed the parameter $q$ of equation~(\ref{ellbarrier}), which is $(1,0.55,0.55,0.707)$ respectively for each model.  This can be incorporated by replacing $\nu \rightarrow q\nu$ in the corresponding formulas.

\subsection{Creation times from Bayes' rule}
Consider the joint probability that a random walk first upcrosses a constant barrier (with $\delta$-intercept between $\delta_{\rm c}$ and $\delta_{\rm c}+{\rm d}\delta_{\rm c}$) between $S$ and $S+{\rm d}S$.  Using Bayes' Theorem, this can be written as
\begin{displaymath}
 P(S,\delta_{\rm c}) {\rm d} S {\rm d} \delta_{\rm c}
 = f(S|\delta_{\rm c})\bar{f}(\delta_{\rm c}) {\rm d} S {\rm d} \delta_{\rm c}
 = c(\delta_{\rm c}|S)\bar{c}(S) {\rm d}\delta_{\rm c}{\rm d}S,
\end{displaymath}
For constant barriers, \cite{percival99} argue that the $\delta_{\rm c}$-prior must be uniform.  First they note that all walks must have a creation event for any barrier, regardless of its height (see Figure~\ref{history}, left panel).  Moreover, for any two equal-sized intervals d$\delta_{\rm c1}$ and d$\delta_{\rm c2}$, the probability that such a creation event exists must be equal.  This is because the steps in the walk are uncorrelated, implying that any point along the walk can be regarded as the starting point of a new walk.  Therefore, the walk is not altered at different values of $\delta_{\rm c}$ and the probability of crossing two different barriers at some point is the same.  In other words, the $\delta_{\rm c}$-prior is given by
\begin{equation}
 \bar{f}(\delta_{c}){\rm d}\delta_{c}
  = \frac{{\rm d}\delta_{c}}{\Delta\delta_{c}},
\label{fdelta}
\end{equation}
where the constant $\Delta\delta_{\rm c}$ is infinite, since $\delta_{\rm c}\in[0,\infty)$.  

Given $\bar{f}(\delta_{\rm c}){\rm d}\delta_{\rm c}$ and $f(S|\delta_{\rm c}){\rm d}S$, one can marginalise the joint distribution in $\delta_{\rm c}$.  This yields 
\begin{equation}
 \label{prs}
 \bar{c}(S){\rm d}S 
  = {\rm d}S\int_{0}^{\infty}{\rm d}\delta_{\rm c}[P(S,\delta_{\rm c})] 
  = \frac{{\rm d}S}{2 {\cal A}\sqrt{S}\Delta\delta_{\rm c}},
\end{equation}
where ${\cal A}=(\sqrt{\pi/2},2,2.08,1.893)$ respectively for each of the four models presented above.  Inserting (\ref{fdelta}) and (\ref{prs}) in Bayes' formula gives 
\begin{equation}
 \label{csfd}
 c(\delta_{\rm c}|S){\rm d} \delta_{\rm c}
 = 2 {\cal A} \sqrt{S} f(S|\delta_{\rm c}){\rm d} \delta_{\rm c}.
\end{equation}
But $f(S|\delta_{\rm c}) = (\nu_{\rm c}/S)f(\nu_{\rm c})$ ($S$ is fixed) 
and $c(\delta_{\rm c}|S)=(2\nu_{\rm c}/\delta_{\rm c})c(\nu_{\rm c})$, 
from which  
\begin{equation}
 c(\nu_{\rm c}){\rm d}\nu_{\rm c} 
 = {\cal A}\sqrt{\nu_{\rm c}}f(\nu_{\rm c}){\rm d}\nu_{\rm c}.
 \label{lucky}
\end{equation}
\end{itemize}
Thus, for constant barriers, there is a remarkably simple relation between the creation distribution $c$ and the first crossing distribution $f$.  

\cite{percival00} argue that equations~(\ref{fdelta}) and~(\ref{lucky}) remain true for all choices of $f$, which in the present context mean, for all barrier shapes.  We will now show why this is incorrect, and also why equation~(\ref{lucky}) nevertheless provides a good approximation to the correct answer.

\subsection{Why it doesn't work in general}\label{linB}


Consider the linear barrier
\begin{equation}
\label{linbarrier}
 B(S,\delta_{\rm c}) =  \delta_{\rm c}\,\Big{[} 1-\beta \Big{(}\frac{S}{\delta_{\rm c}^{2}}\Big{)}\Big{]}.
\end{equation} 
This is special case of the barrier in equation~(\ref{ellbarrier}) with $\gamma=1$ (we have suppressed the parameter $q$).  Notice that we have deliberately replaced $\beta \rightarrow -\beta$ in the above expression.  In this work we will only consider the $\beta>0$ case to avoid barriers that intersect.  Moreover, if $\beta>0$, all walks are guaranteed to cross.  The exact solution is known, and it is given by
\begin{equation}
f(\nu){\rm d}\nu = \sqrt{\frac{\nu}{2\pi}}{\rm e}^{-\frac{\nu}{2}(1-\frac{\beta}{\nu})^2}\frac{{\rm d}\nu}{\nu}\,,
\label{fnulin}
\end{equation}
where $\nu=\delta_{\rm c}^2/S$ (fixed $S$) \citep{schroedinger15,sheth98}.  If equation~(\ref{fdelta}) holds, then the creation time distribution should be given by 
\begin{equation}
\label{cnulin}
c(\nu_{\rm c}){\rm d}\nu_{\rm c}={\cal A}\sqrt{\nu_{\rm c}}f(\nu_{\rm c}){\rm d}\nu_{\rm c}, {\rm \,\,\,\,where\,\,}{\cal A}=\sqrt{\frac{\pi}{2}}{\rm e}^{-\frac{\beta}{2}},
\end{equation}
and $\nu_{\rm c}=\delta_{\rm c}^2/S$ (fixed $\delta_{\rm c}$).  

\begin{figure}
 \centering
 \includegraphics[width=\hsize]{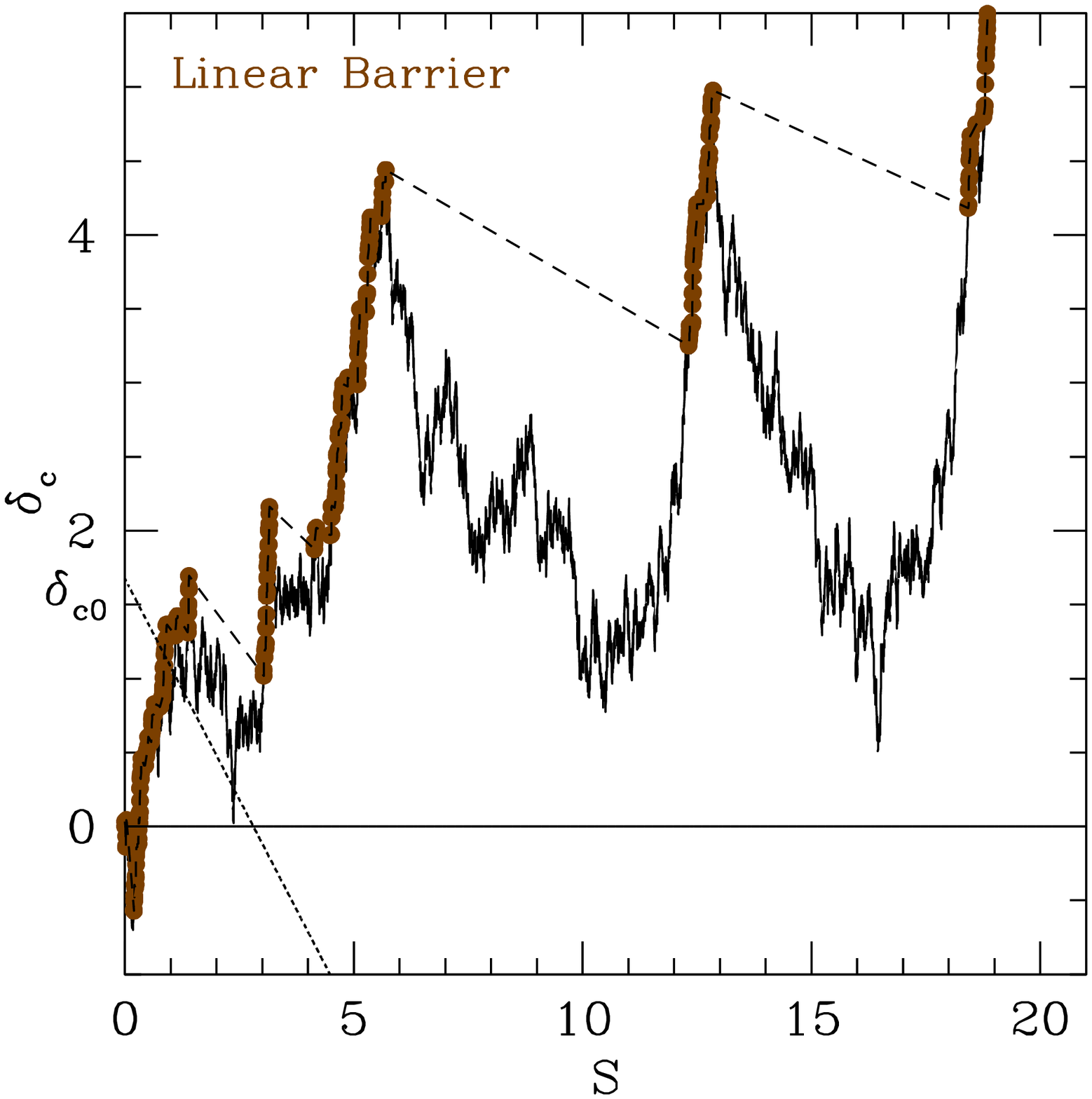}
 \caption{The mass history associated with a random walk (jagged line) and linear barriers.  The brown filled circles on the random walk denote the creation events.  Notice that the barriers become steeper as $\delta_{\rm c}$ increases, and the separation between any two barriers increases with increasing $S$.  For reference, the dotted line in the lower left represents the barrier associated with $\delta_{\rm c}=\delta_{\rm c0}$.}
 \label{linhist}
\end{figure}

\begin{figure}
 \centering
 \includegraphics[width=\hsize]{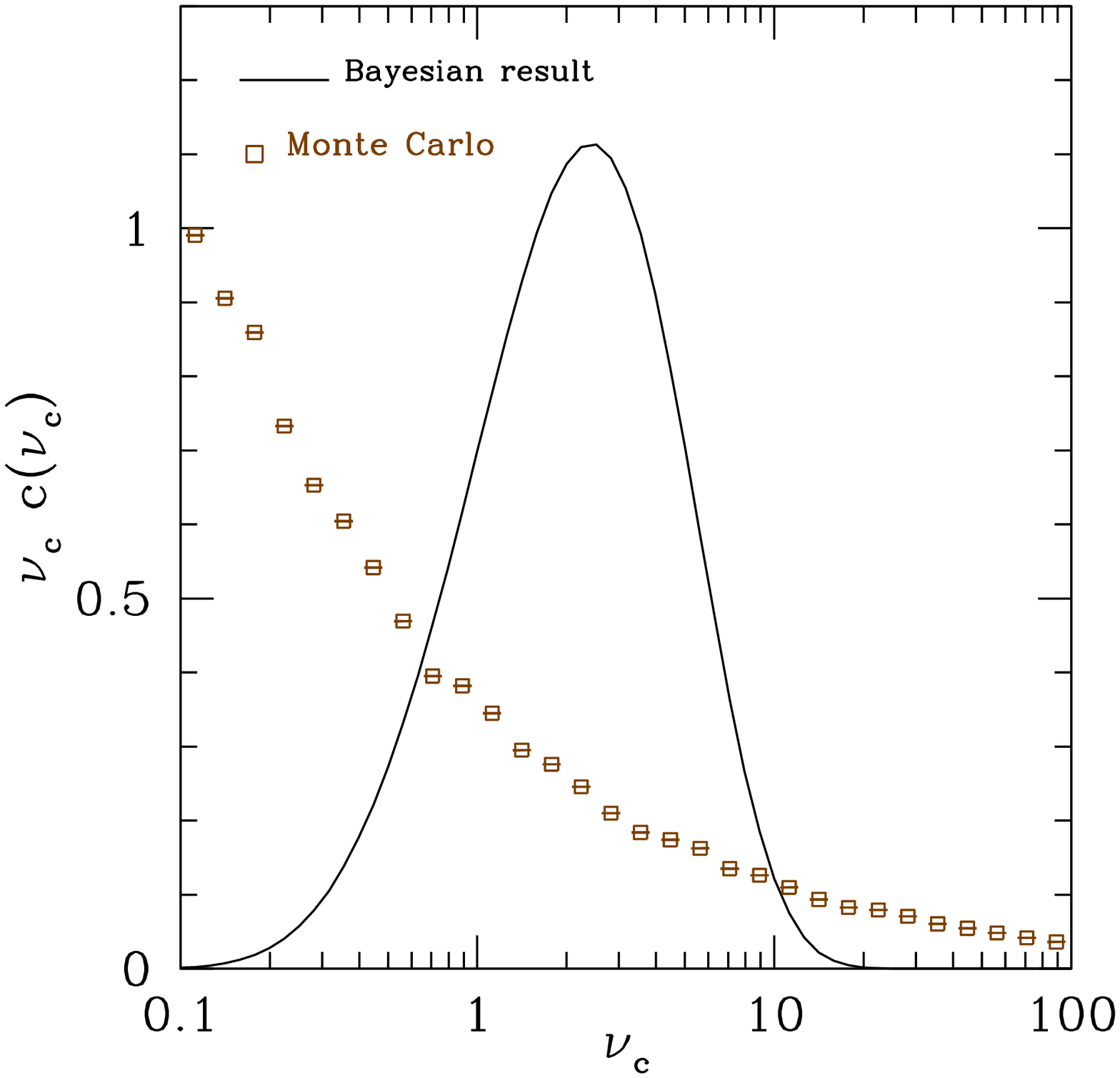}
 \caption{The creation time distribution associated with linear barriers in self-similar form.  The variable $\nu_{\rm c}$ denotes $\delta_{\rm c}^{2}/S$ at fixed mass.  The squares show the distribution measured from an ensemble of random walks with linear barriers.  The solid curves show the associated predictions -- assuming that equation~(\ref{fdelta}) applies (the `Bayesian' result).  The discrepancy between the theory prediction and the data indicates that care must be taken when using the result in equation~(\ref{cvfv}).}
 \label{linbayes}
\end{figure}

To test this, we performed a Monte Carlo simulation of the mass histories associated with random walks where linear barriers (with $\beta=1$) are used to select creation events.  Figure~\ref{linhist} shows one sample mass history of this ensemble.  Compare this with Figure~\ref{history}, which shows the same process, but with constant and square-root barriers.  The jagged line is a random walk, and the brown solid circles are the associated history.  The long dashed lines denote jumps in the history.  For completeness we have included the linear barrier with $\delta_{\rm c}=\delta_{\rm c0}$, depicted as a dotted line in the lower left.  Figure~\ref{linbayes} shows our Monte Carlo data (brown open squares).  The solid black curve is the prediction in equation~(\ref{cnulin}).  This disagreement invalidates the claim that $c(\nu_{\rm c})$d$\nu_{\rm c}$ is {\it always} proportional to $\sqrt{\nu_{\rm c}}f(\nu_{\rm c})$d$\nu_{\rm c}$.  

Before moving on, note that the Monte Carlo data in Figure~\ref{cnulin} follow a smooth curve that is quite different from that associated with constant or square-root barriers (Figure~\ref{times}).  The main difference is that the latter peak at some intermediate value of $\nu_{\rm c}$, whereas the former decreases monotonically with $\nu_{\rm c}$.  This is because it is unlikely for a random walk to upcross a constant barrier (or a square-root barrier) after a few steps.  In other words, creation events with $S \ll \delta_{\rm c}^2$ (i.e., $\nu_{\rm c} \ll 1$) are very unlikely.  Similarly, it is unlikely that a walk survives for many steps without being absorbed by a constant barrier.  That is, creation events with $S \gg \delta_{\rm c}^2$ (i.e., $\nu_{\rm c} \gg 1$) are unlikely.  For the linear barrier, the ensemble of creation events is dominated by points with small $\delta_{\rm c}$ (i.e., small $\nu_{\rm c}$).  This is because the linear barrier becomes steeper as $\delta_{\rm c}$ decreases.  Since the height of a linear barrier decreases with $S$, it becomes easier for walks to upcross a barrier as $\delta_{\rm c}$ decreases.  This even allows for cases where creation events are selected from points along a walk with $\delta < 0$.  Such cases would be impossible for the constant and the square-root barrier models.  

In principle, the creation time distribution with linear barriers can be computed analytically, without using the Bayesian approach presented here \cite[e.g.,][]{karlin75}.  Since the mass function associated with linear barriers does not resemble halo abundances in N-body simulations, we do not pursue this any further \cite[but see][for interesting applications of this barrier]{sheth98}.

\subsection{Why it is a useful approximation in practice}
If equation~(\ref{lucky}) is incorrect in general, then why then did it work so well in the main text?  The first step is to recognize that, because of the property highlighted by equation~(\ref{whyitworks}), a uniform distribution in $\delta_{\rm c}$ is appropriate for the family of square-root barriers of interest to us (equation~\ref{sqrtbarrier}), and so, for square-root barriers, equation~(\ref{lucky}) is exact.  
Our Monte Carlo simulations shown in the main text confirm that this is indeed the case.  

The second step is to note that, for barriers with general $\gamma$, the difference between two barriers carries additional factors of $\delta_{\rm c}$ (e.g., equation~\ref{ellbarrier}).  As a result, assumption~(\ref{fdelta}) is no longer valid.  
E.g., for linear barriers,  
\begin{equation}
 B(S,\delta_{\rm c2})-B(S,\delta_{\rm c1}) = 
 \delta_{\rm c2}-\delta_{\rm c1}-\beta \Big{(}\frac{S}{\delta_{\rm c2}}-\frac{S}{\delta_{\rm c1}}\Big{)} \neq \delta_{\rm c2}-\delta_{\rm c1}.
\end{equation}
This property makes the linear barrier considerably different from the constant and square-root barriers.  Figure~\ref{linbayes} shows that the distance between any two linear barriers increases with increasing $S$ (compare with Figure~\ref{history}), so the $\delta_{\rm c}$-prior is not uniform. 

It is important to emphasize that the central conclusion of this and the previous subsection -- that the assumptions behind equation~(\ref{cvfv}) do not hold in general -- do not depend on the fact that the linear barrier decreases in height with $S$.  E.g., barriers of the form given by equation~(\ref{ellbarrier}) with $\beta>0$ and $0<\gamma<1/2$ increase with $S$.  However, 
\begin{equation}
 B(S,\delta_{\rm c2})-B(S,\delta_{\rm c1})\neq \delta_{\rm c2}-\delta_{\rm c1};
\end{equation}
the separation between any two barriers increases with $S$, so the $\delta_{\rm c}$-prior is not uniform in this case either.  For these barriers too, equation~~(\ref{lucky}) is incorrect.  Nevertheless, for $\gamma$ close to 0 or 1/2, equation~(\ref{lucky}) should provide a reasonable approximation.  This is the fundamental reason why barriers with $\gamma=0.6$, or of the form required to give the \cite{sheth99} formula as the first crossing distribution, are likely to have creation time distributions which are well approximated by equation~(\ref{lucky}).

\subsection{Conditional formulae}\label{condapp}

The conditional mass function and creation times distribution can be obtained from the unconditional crossing distribution of the barrier ${\cal B}$ in equation~(\ref{dbarrier}).  The natural variables in this case are $s/S_0$ and $\eta_\beta=a_{\rm c}/\sqrt{S_0}$ (where $a_{\rm c}\equiv\delta_1-\delta_0-\beta\sqrt{S_0}$).  Explicity, the crossing distribution can be written as
\begin{equation}
\label{fg}
f(s/S_{0}|\eta_{\beta}){\rm d}(s/S_{0}) = g(s/S_0,\eta_{\beta})  \frac{{\rm d} (s/S_{0})}{s/S_{0}+1}.
\end{equation} 
where two forms of $g$ are given below \citep{breiman66,sheth02}.
\begin{itemize}
\item Square-root barrier (exact):
\begin{equation}
g(s/S_0,\eta_{\beta})= \sum_{\{\lambda\}} {\rm e}^{{\eta_{\beta}}^2/4} \frac{ D_{\lambda}(\eta_{\beta})l_{\lambda}(-\beta)}{(s/S_{0}+1)^{\lambda/2} }.
\end{equation}
\item Square-root barrier (series approximation):
\begin{eqnarray}
g(s/S_0,\eta_{\beta})=\frac{ \Big{|} \eta_{\beta} + \beta \sqrt{s/S_{0}+1} \big{[} 1+ \alpha_{(s/S_{0})} \big{]} \Big{|} } {\sqrt{ 2\pi s/S_{0} } } 
\nonumber \\ \times \exp{ \Big{\{} -\frac{ (\eta_{\beta}+\beta \sqrt{s/S_{0}+1})^2  }{ 2 s/S_{0} }  \Big{\}} } \frac{s/S_0+1}{s/S_0},
\end{eqnarray}
where
\begin{displaymath}
\alpha_{(s/S_0)} = \sum^{5}_{n=1} \frac{\alpha_{n}}{s/S_0+1}, \,\,\,\alpha_{\,1}=-\frac{1}{2}, \,\,\, {\rm and}\,\,\, \alpha_{n}=(1-\frac{3}{2n})\alpha_{n-1}.
\end{displaymath}

\item Bayes' rule:

The joint distribution of $s/S_0$ and $\eta_\beta$ is given by
\begin{eqnarray}
\label{cjoint}
P(s/S_0,\eta_\beta){\rm d}(s/S_0){\rm d}\eta_\beta=f(s/S_0|\eta_\beta)\bar{f}(\eta_\beta){\rm d}(s/S_0){\rm d}\eta_\beta \nonumber \\ = c(\eta_\beta|s/S_0)\bar{c}(s/S_0){\rm d}(s/S_0){\rm d}\eta_\beta .
\end{eqnarray} 
Following equation~(\ref{fdelta}) for $\delta_1$ and $\delta_0$ and using the fact that $S_0$ is fixed, it can be shown that the $\eta_\beta$-prior is uniform:
\begin{equation}
 \label{etapr}
 \bar{f}(\eta_\beta){\rm d}\eta_\beta = \frac{{\rm d}\eta_\beta}{\Delta \eta}
\end{equation}
where $\Delta \eta$ is an infinite constant.  Marginalising over the joint distribution in $\eta_\beta$ we obtain
\begin{equation}
\label{spr}
 \bar{c}(s/S_0){\rm d}(s/S_0)  \frac{G}{s/S_0+1}\frac{{\rm d}(s/S_0)}{\Delta \eta},
\end{equation}
where
\begin{equation}
 G(s/S_0) = \int^{\infty}_0 {\rm d}\eta_\beta\, g(s/S_0,\eta_\beta).
\end{equation}
Inserting equations~(\ref{fg}), (\ref{etapr}) and (\ref{spr}) in Bayes' rule, we obtain
\begin{equation}
 \label{ceta}
 c(\eta_{\beta}|s/S_{0}){\rm d}\eta_{\beta} = \frac{g}{G} {\rm d} \eta_{\beta}.
\end{equation}
Comparing~(\ref{fg}) to (\ref{ceta}), we find that 
\begin{equation}
 c(\eta_\beta|s/S_{0}){\rm d}\eta_\beta = 
 {\cal A}_{(s/S_0)} f(s/S_0|\eta_\beta){\rm d} \eta_{\beta},
\end{equation}
where
\begin{equation}
 {\cal A}_{(s/S_0)}= (s/S_0+1)/G.
\end{equation}
This result is confirmed by our Monte Carlo simulations (Figure~\ref{ctimes}).

\end{itemize}

\section{Coagulation Theory}\label{smolapp}

\subsection{White-noise initial conditions}

The creation rate in the discrete Smoluchowski equation is 
\begin{equation}
 C(m,t) = \sum_{m'=1}^{m-1}\frac{K(m',m-m';t)}{2}n(m'|t)n(m-m'|t).
\end{equation}
If the kernel $K$ is additive, the solution approaches the white-noise Press-Schechter mass function in the continuum limit (large $m$ and small $\delta_{\rm c}$).  Moreover, \cite{sheth97} showed that the above equation can be written as
\begin{equation}
C(m,t)=\bar{n}\,n(m|t)\frac{m-1}{1+\delta_{\rm c}} \Big{|}\frac{{\rm d}\delta_{\rm c}}{{\rm d}t}\Big{|}\sum_{m'=1}^{m-1}p_{m'|m-m'},
\end{equation}
where $\bar{n}$ is the mean number-density of particles.  If we picture haloes as a collection of particles held together by $(m-1)$ non-intersecting bonds (branched polymers), $p_{m'|m-m'}$ gives the probability of obtaining an $m'$-halo and and $(m-m')$-halo by deleting one random bond in the $m$-halo.  The sum over a normalised probability is trivial, so the creation term is simply
\begin{equation}
C(m,t)=\bar{n}\,n(m|t)\frac{m-1}{1+\delta_{\rm c}} \Big{|}\frac{{\rm d}\delta_{\rm c}}{{\rm d}t}\Big{|}.
\end{equation}
In the continuum limit, this becomes
\begin{equation}
\label{mna}
C(m,t)=\bar{\rho}\,m\,n(m|t) \Big{|}\frac{{\rm d}\delta_{\rm c}}{{\rm d}t}\Big{|}.
\end{equation}

Similarly, in the conditional case, 
\begin{equation}
 C(m,t|M,T) = \bar{n}\,N(m|t,M,T)\frac{m-1}{1+\delta_{\rm c}} 
 \Big{|}\frac{{\rm d}\delta_{\rm c}}{{\rm d}t}\Big{|}
\end{equation}
\citep{sheth03}.  
In the continuum limit, this becomes
\begin{equation}
\label{mNa}
 C(m,t\,|M,T) = \bar{\rho}\,m\,N(m|t,M,T)\,
 \Big{|} \frac{{\rm d}\delta_{\rm c}}{{\rm d}t} \Big{|}.
\end{equation}

Notice that the two rates (equations~\ref{mna} and \ref{mNa}) are related by the following consistency condition:
\begin{equation}
\int^{\infty}_{m}\,{\rm d}M \,C(m,t\,|M,T)\,n(M|T)=C(m,t).
\end{equation}
In other words, the unconditional rate is recovered from the conditional rate by multiplying by the number density of $M$-haloes at $T$ and integrating over all possible $M>m$.  This consistency relation is true in general, not just when the initial conditions are white-noise.

\subsection{The time-normalised creation rate}\label{Cc}

In this appendix, we show that creation rate $C(m,t)$ (equation~\ref{mn}, Section~4) is related to the creation time distribution $c(t|m)$ (equation~\ref{cvfv}, Section~3) in a simple way.  First, notice that $c(t|m)$ can be written in terms of $\nu_{\rm c}=\delta^2_{\rm c}/S$.  That is,
\begin{equation}
c(t|m)=c(\delta_{\rm c}|S)\Big{|}\frac{{\rm d}\delta_{\rm c}}{{\rm d}t}\Big{|}=c(\nu_{\rm c})\Big{|}\frac{{\rm d}\nu_{\rm c}}{{\rm d}\delta_{\rm c}}\Big{|}\Big{|}\frac{{\rm d}\delta_{\rm c}}{{\rm d}t}\Big{|}=\frac{2\nu_{\rm c}}{\delta_{\rm c}}c(\nu_{\rm c})\Big{|}\frac{{\rm d}\delta_{\rm c}}{{\rm d}t}\Big{|},
\end{equation}
where we have used the fact that $|{\rm d}\ln \nu_{\rm c}/{\rm d}\ln \delta_{\rm c}|=2$. 

 Now, following equation~(\ref{cvfv}), the above expression can be written as
\begin{equation}
c(t|m)=\frac{2\nu_{\rm c}}{\delta_{\rm c}}{\cal A}\sqrt{\nu_{\rm c}}f(\nu_{\rm c})\Big{|}\frac{{\rm d}\delta_{\rm c}}{{\rm d}t}\Big{|}.
\label{den}
\end{equation}
On the other hand,
\begin{equation}
\frac{mn(m|t)}{\bar{\rho}}=f(m|t)=f(S|\delta_{\rm c})\Big{|}\frac{{\rm d}S}{{\rm d}m}\Big{|}=\frac{\nu}{S}f(\nu)\Big{|}\frac{{\rm d}S}{{\rm d}m}\Big{|},
\end{equation}
where we have used the fact that $|{\rm d}\ln \nu/{\rm d}\ln S|=1$. 

Our prescription for the creation rate (equation~\ref{mn}) is
\begin{equation}
C(m,t)=\bar{\rho}^2\frac{mn(m|t)}{\bar{\rho}}\Big{|}\frac{{\rm d}\delta_{\rm c}}{{\rm d}t}\Big{|}=\bar{\rho}^2 \frac{\nu}{S}f(\nu)\Big{|}\frac{{\rm d}S}{{\rm d}m}\Big{|}  \Big{|}\frac{{\rm d}\delta_{\rm c}}{{\rm d}t}\Big{|}.
\label{num}
\end{equation}

Taking the ratio of (\ref{num}) and (\ref{den}), we obtain
\begin{equation}
\frac{C(m,t)}{c(t|m)}=\frac{ \bar{\rho}^2 \frac{\nu}{S}f(\nu)\Big{|}\frac{{\rm d}S}{{\rm d}m}\Big{|}  \Big{|}\frac{{\rm d}\delta_{\rm c}}{{\rm d}t}\Big{|}   }{  \frac{2\nu_{\rm c}}{\delta_{\rm c}}{\cal A}\sqrt{\nu_{\rm c}}f(\nu_{\rm c})\Big{|}\frac{{\rm d}\delta_{\rm c}}{{\rm d}t}\Big{|}}=\frac{\bar{\rho}^2}{2{\cal A \sqrt{S}}}\Big{|}\frac{{\rm d}S}{{\rm d}m}\Big{|}\equiv g(m),
\end{equation}
where we have dropped the distinction in notation between $\nu$ and $\nu_{\rm c}$.  Simply put,
\begin{equation}
C(m,t)=g(m)c(t|m).
\end{equation}
Integrating both sides with respect to time,
\begin{equation}
\int C(m,t) {\rm d}t= \int g(m)c(t|m) {\rm d}t = g(m) \int c(t|m) {\rm d}t= g(m),
\end{equation}
since $c(t|m)$ is, by definition, a time-normalised distribution.
Thus, 
\begin{equation}
c(t|m)=\frac{C(m,t)}{g(m)}=\frac{C(m,t)}{\int C(m,t) {\rm d}t}.
\end{equation}
In other words, $c(t|m)$ can be obtained from $C(m,t)$ by normalising the latter in time.

\label{lastpage}
\end{document}